\documentclass{aa}  

\usepackage{graphicx}
\usepackage{txfonts}
\usepackage{hyperref}
\hypersetup{
    colorlinks=true,
    citecolor=blue,
    linkcolor=blue,
    filecolor=magenta,      
    urlcolor=blue,
    pdftitle={Overleaf Example},
    pdfpagemode=FullScreen,
    }

\begin{document} 

   \title{Photometric Binaries in 14 Magellanic Cloud Star Clusters}

   \author{Anjana Mohandasan
          \inst{1},
          Antonino P. Milone
          \inst{1,}
          \inst{2},
          Giacomo Cordoni
          \inst{2,}
          \inst{3},
          Emanuele Dondoglio
          \inst{1},
          Edoardo P. Lagioia
          \inst{1,}\inst{4},
           Maria Vittoria Legnardi
          \inst{1},
          Tuila Ziliotto
          \inst{1},
          Sohee Jang
          \inst{5},
          Anna F. Marino
          \inst{2,}
          \inst{6},
          \and 
          Mar\'ilia Carlos
          \inst{7}
          }

   \institute{Dipartimento di Fisica e Astronomia "Galileo Galilei", Università Degli Studi Di Padova, Vicolo dell'Osservatorio 3, Padova, IT-35122\\
              \email{anjana.mohandasan@studenti.unipd.it}
         \and
             Istituto Nazionale di Astrofisica – Osservatorio Astronomico di Padova, Vicolo dell’Osservatorio 5, 35122 Padova, Italy
             \and
             Research School of Astronomy and Astrophysics, Australian National University, Canberra ACT 2601, Australia
             \and
             South-Western Institute for Astronomy Research, Yunnan University, Kunming, 650500 P.R.China
             \and
             Center for Galaxy Evolution Research and Department of Astronomy, Yonsei University, Seoul 03722, Korea
             \and
             Istituto Nazionale di Astrofisica – Osservatorio Astronomico di Arcetri, Largo Enrico Fermi, 5, 50125 Firenze, Italy
             \and 
             Department of Physics and Astronomy, Uppsala University, Box 516, SE-751 20 Uppsala, Sweden
             }

\large
\abstract{ Binary stars play a major role in determining the dynamic evolution of star clusters.
We used images collected with the Hubble Space Telescope to study fourteen Magellanic Clouds star clusters that span an age interval between $\sim 0.6$ and  $2.1$ Gyr and masses of $10^{4}-10^{5}$ M$_{\odot}$. 
We estimated the fraction of binary systems composed of two main-sequence stars and the fraction of candidate blue-straggler stars (BSSs).
Moreover, we derived the structural parameters of the cluster, including the core radius, the central density, the mass function, and the total mass. 
We find that the fraction of binaries with a mass ratio larger than 0.7 ranges from $\sim$7\%, in NGC\,1846, to $\sim$20\%, in NGC\,2108.
The radial and luminosity distribution can change from one cluster to another.
However, when we combine the results from all the clusters, we find that binaries follow a flat radial trend and no significant correlation with the mass of the primary star.  We find no evidence for a relation between the fractions of binaries and BSSs.
We combined the results on binaries in the studied Magellanic Cloud clusters with those obtained for 67 Galactic globular clusters and 78 open clusters.  
We detect a significant  anti-correlation between the binary fraction in the core and the mass of the host cluster. However, star clusters with similar masses exhibit a wide range of binary fractions. Conversely, there is no evidence of a correlation between the fraction of binaries and either the cluster age or the dynamic age.} 
 
   \titlerunning{Photometric binaries}
   \authorrunning{A. Mohandasan}
   \maketitle
   
%
\nolinenumbers
\section{Introduction} \label{sec:intro}

\par Being dense environments, star clusters are home to a variety of intriguing objects. Among them, binaries hold a position that is both fundamental and significant. Both open and globular clusters (GCs) host a substantial number of binaries, many of them being primordial \citep{hut1992a}. Binaries have a major role to play in the ever-ensuing dynamics of the clusters and are pivotal in determining their various parameters such as age, luminosity function, radius, and mass.

\par Binaries have higher interaction cross-sections compared to single stars. Through the frequent interactions with other objects in the cluster, they form an efficient mechanism of kinetic energy distribution. This binary-burning phase of the cluster is the longest phase of cluster evolution in which most of the observed clusters are expected to be \citep{heggie1975}. The elastic scattering interactions of binaries are effective in withstanding the impeding gravitational collapse in the center, thereby, defining the dense cluster core \citep{heggie2003}. The high rate of scattering transforms the dense cluster neighborhood into an ideal cauldron for the production of exotic objects. In recent decades, Blue Straggler Stars \citep[BSSs,][]{piotto2004}, Cataclysmic Variables \citep[CVs,][]{cool1995}, Low-Mass X-ray Binaries \citep[LMXBs,][]{kim2006}, and Millisecond Pulsars \citep[MSPs,][]{peter2002} have been detected in close binary systems in star clusters. And, their evolution is better understood against the backdrop of binary fraction in the cluster.  

\par The evolution of binaries in a cluster largely depends on two factors; i) the secular binary stellar evolution and ii) their interaction with other cluster members. Secular binary evolution can depend on the mass and mass ratio of the companion stars. The interaction rate in clusters can be significant enough to cause the destruction of soft binaries even in low-density stellar systems. Hard binaries, on the other hand, become even harder with interactions \citep{heggie1975}. These interactions can guide the binary stellar evolution by altering the orbital parameters, assisting the escape and exchange of the binary members. The binaries with kinetic energy comparable to that of a typical cluster member are the main variables in the binary fraction. Their amount and influence on cluster evolution can significantly depend on the environment.  
By exploring the binary fractions in star clusters belonging to diverse environments, in terms of their different stellar and cluster parameters, we can peek at the dynamics underway.

\par Among the various techniques in literature to identify and characterize binary systems, the method based on the color-magnitude diagram (CMD) provides an efficient approach for constraining the fraction of unresolved main sequence (MS) binaries \citep{rub1997}. In the CMD of a star cluster, the MS-MS binaries populate the region on the red side of the MS fiducial line (MSFL).
The two other prominent detection methods for binaries are (1) radial velocity variability analysis, which is employed to analyse binaries comprising a massive companion \citep{latham1996}, and (2) the method based on the stellar photometric variability \citep{mateo1996}. These methods are biased toward the brightest systems or the binaries with short and eccentric orbits. CMD analysis not only overcomes these limitations but is also statistically robust. A limited amount of observational time in 2 filters alone is sufficient to examine the multitude of cluster members belonging to a large parameter space \citep{hut1992a}. 
  
\par Photometric errors, differential reddening, and spatial dependent variations of the photometric zero points are among the major challenges for an accurate determination of photometric binaries. Field stars that sneak into the cluster region of the field of view (FoV) pose another hurdle. Hence, this approach necessitates high-precision photometry, astrometry, and high-resolution images which are achievable in the era of the Hubble Space Telescope (HST) and advanced data reduction software.

\par Pioneering studies on photometric binaries can be seen in the work of \cite{bolte1992}, \cite{aparacio1991}, and \cite{romani1991}. The first investigation of a large sample of clusters was done by \cite{sollima2007}, who studied the binaries in 13 low-density Galactic GCs and derived the binary fractions in the cluster cores. An extensive survey of photometric binaries in 67 Galactic GCs was conducted by \cite{milone2012} and \cite{milone2016}, using the images collected with the Advanced Camera for Survey (ACS) aboard HST.
They analysed the binary fraction as a function of various cluster and stellar parameters and found that binaries in the GCs are typically more centrally concentrated than single stars, while the fraction of binaries in the cluster core anti-correlates with the stellar mass. 

The main limitation of these works is that the Galactic GCs share similar ages, with most of them being older than $\sim$10 Gyr and massive \citep{dotter2010}. Hence, it is challenging to investigate the dependence of binary fractions on cluster age and study the binary systems where one component is more massive than $\sim$0.8$M_{\odot}$.
On the contrary, the investigation of binaries in Galactic open clusters is limited to low-mass clusters with masses smaller than $\sim$10$^{4}$ solar masses \citep[e.g.][]{cordoni2023}.
 
 The present work surpasses these limitations on mass and age by 
analysing fourteen star clusters in the Large and Small Magellanic Clouds (LMC and SMC) spanning an age interval between $\sim$0.6 and 2.1 Gyrs and masses between 10$^4$ and $10^{5}$ solar masses. 
\par The paper is structured as follows. Section\,\ref{sec:style} is dedicated to the details of observations, data sources, and data reduction techniques. In Section\,\ref{sec:parametri} we derive the structural parameters of the star cluster, whereas Section\,\ref{sec:method} discusses the method for deriving the binary fractions. 
Section\,\ref{sec:floats} presents the results, while a summary of the paper is provided in Section\,\ref{sec:conclusions}. 

\section{Data source and data reduction } \label{sec:style}
\begin{table*}
\caption{Information on the HST images used in this paper. For each cluster, we provide the available filters, the camera, the number of images, and the corresponding exposure times. We also list the date of the observation, the program, and the name of the principal investigator.}
\label{table:1}
\centering
    \begin{tabular*}{0.9\textwidth}{@{\extracolsep{\fill}} c c c c c c c}
    \hline
    \hline
    Cluster ID & Filter & Instrument &  N$\times$Exposure Time & Date & Program & PI\\
    \\
    \hline
ESO057SC075 & F435W & WFC/ACS & 55s $+$ 2$\times$340s & Nov 17 2006 & 10595 & P. Goudfrooij \\
 & F814W & WFC/ACS & 15s $+$ 2$\times$340s & Nov 17 2006 & 10595 & P. Goudfrooij\\
ESO057SC030 & F475W & UVIS/WFC3 & 120s $+$ 600s $+$ 720s & August 16 2012 & 12257 & L. Girardi \\
 & F814W & UVIS/WFC3 & 30s $+$ 2$\times$700s & August 16 2012 & 12257 & L. Girardi \\
 KMHK316 & F475W & WFC/ACS & 2$\times$665s & June 10 2016 & 14204 & A. P. Milone\\
 & F814W & WFC/ACS & 42s $+$ 533s & June 10 2016 & 14204 & A. P. Milone\\
NGC1651 & F475W & UVIS/WFC3 & 120s $+$ 600s $+$ 720s & Oct 16 2011 & 12257 & L. Girardi \\
  & F814W & UVIS/WFC3 & 30s $+$ 2$\times$700s & Oct 16 2011 & 12257 & L. Girardi \\
NGC1718 & F475W & UVIS/WFC3 & 120s $+$ 600s $+$ 720s & Dec 02 2011 & 12257 & L. Girardi  \\
  & F814W & UVIS/WFC3 & 30s $+$ 2$\times$700s & Dec 02 2011 & 12257 & L. Girardi  \\
NGC1751 & F435W & WFC/ACS & 90s $+$ 2$\times$340s & 18 Oct 2006 & 10595 & P. Goudfrooij \\
  & F814W & WFC/ACS & 8s $+$ 200s $+$ 2$\times$340s & 18 Oct 2006 & 10595 & P. Goudfrooij \\
NGC1783 & F435W & WFC/ACS & 90s $+$ 2$\times$340s & Jan 14 2006 & 10595 & P. Goudfrooij \\
  & F814W & WFC/ACS & 8s $+$ 170s $+$ 2$\times$340s & Jan 14 2006 & 10595 & P. Goudfrooij \\
NGC1806 & F435W & WFC/ACS & 90s $+$ 2$\times$340s & Sep 29 2005 & 10595 & P. Goudfrooij \\
  & F814W & WFC/ACS & 8s $+$ 200s $+$ 2$\times$340s & Sep 29 2005 & 10595 & P. Goudfrooij \\
NGC1846 & F435W & WFC/ACS & 90s $+$ 2$\times$340s & Jan 12 2006 & 10595 & P. Goudfrooij \\
  & F814W & WFC/ACS & 8s $+$ 2$\times$340s & Jan 12 2006 & 10595 & P. Goudfrooij \\
  & F814W & WFC/ACS & 200s & Oct 08 2003 & 9891 & G. Gilmore \\
NGC1868 & F336W & UVIS/WFC3 & 2$\times$831s $+$ 830s & Dec 22 2016 & 14710 & A. P. Milone \\
 & F814W & UVIS/WFC3 & 90s $+$ 666s & Dec 22 2016 & 14710 & A. P. Milone \\
NGC1872 & F555W & WFC/ACS & 115s & Sep 21 2003 & 9891 & G.Gilmore \\
 & F814W & WFC/ACS & 90s & Sep 21 2003 & 9891 & G.Gilmore \\
NGC2108 & F435W & WFC/ACS & 90s $+$ 2$\times$340s & Aug 22 2006 & 10595 & P. Goudfrooij \\
 & F814W & WFC/ACS & 8s $+$ 2$\times$340s & Aug 22 2006 & 10595 & P. Goudfrooij \\
 & F814W & WFC/ACS & 170s & Aug 16 2003 & 9891 & G. Gilmore \\
NGC2203 & F475W & UVIS/WFC3 & 120s $+$ 2$\times$700s & Oct 08 2011 & 12257 & L. Girardi \\
 & F814W & UVIS/WFC3 & 30s $+$ 550s $+$ 2$\times$700s & Oct 08 2011 & 12257 & L. Girardi \\
 NGC2213 & F475W & UVIS/WFC3 & 120s $+$ 600s $+$ 720s & Nov 29 2011 & 12257 & L. Girardi \\
& F814W & UVIS/WFC3 & 30s $+$ 2$\times$700s & Nov 29 2011 & 12257 & L. Girardi \\
\hline
\hline
    \end{tabular*}
\end{table*}

\begin{figure}[ht]
\centering
\includegraphics[width=9cm]{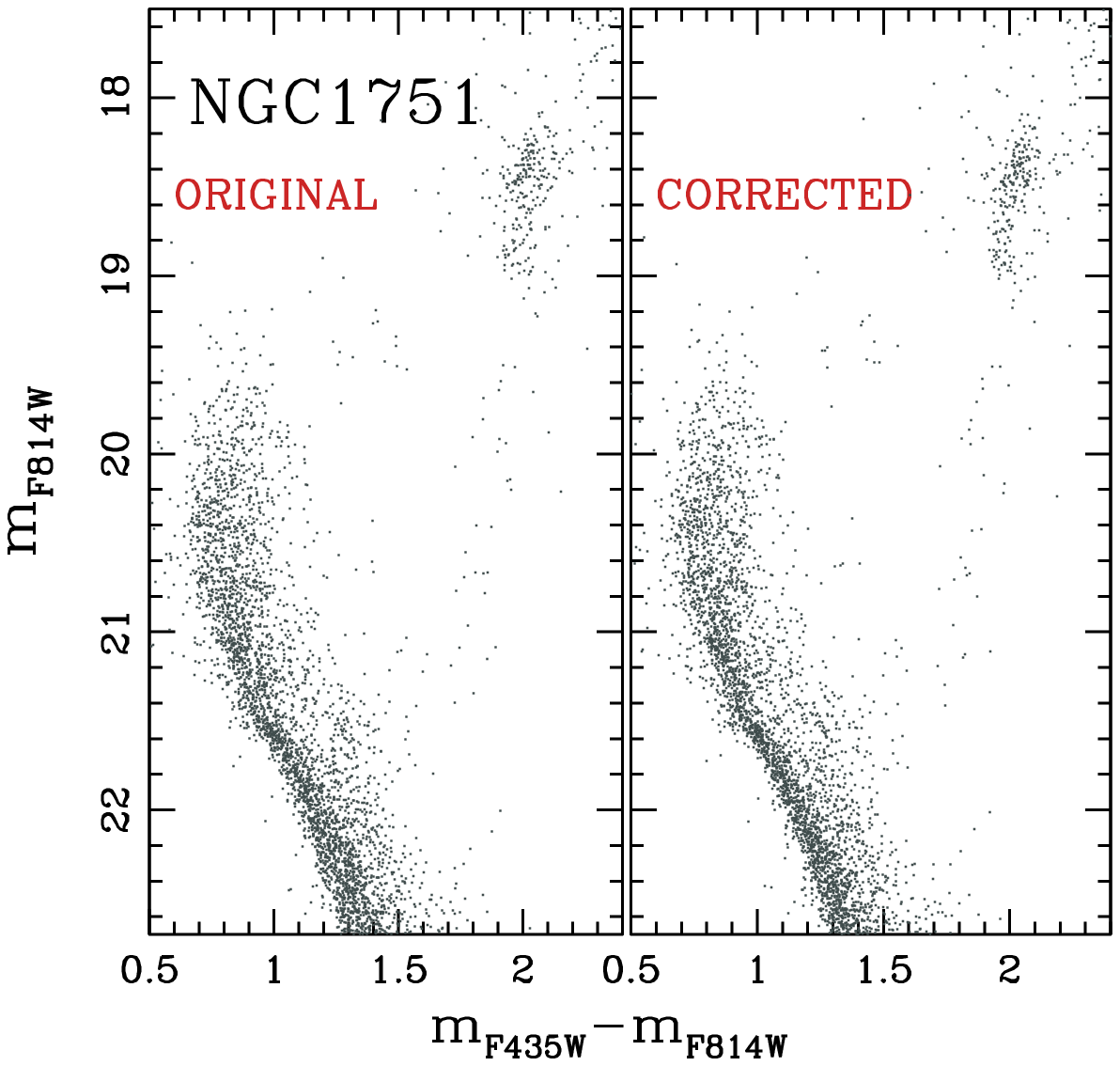}
\caption{Differential reddening correction. The figure comparison of the CMD of the star cluster NGC\,1751 before ({\it{left panel}}) and after the correction for differential reddening ({\it{right panel}}).}
\label{fig:dr}
\end{figure}

\begin{figure*}
\includegraphics[width=18cm]{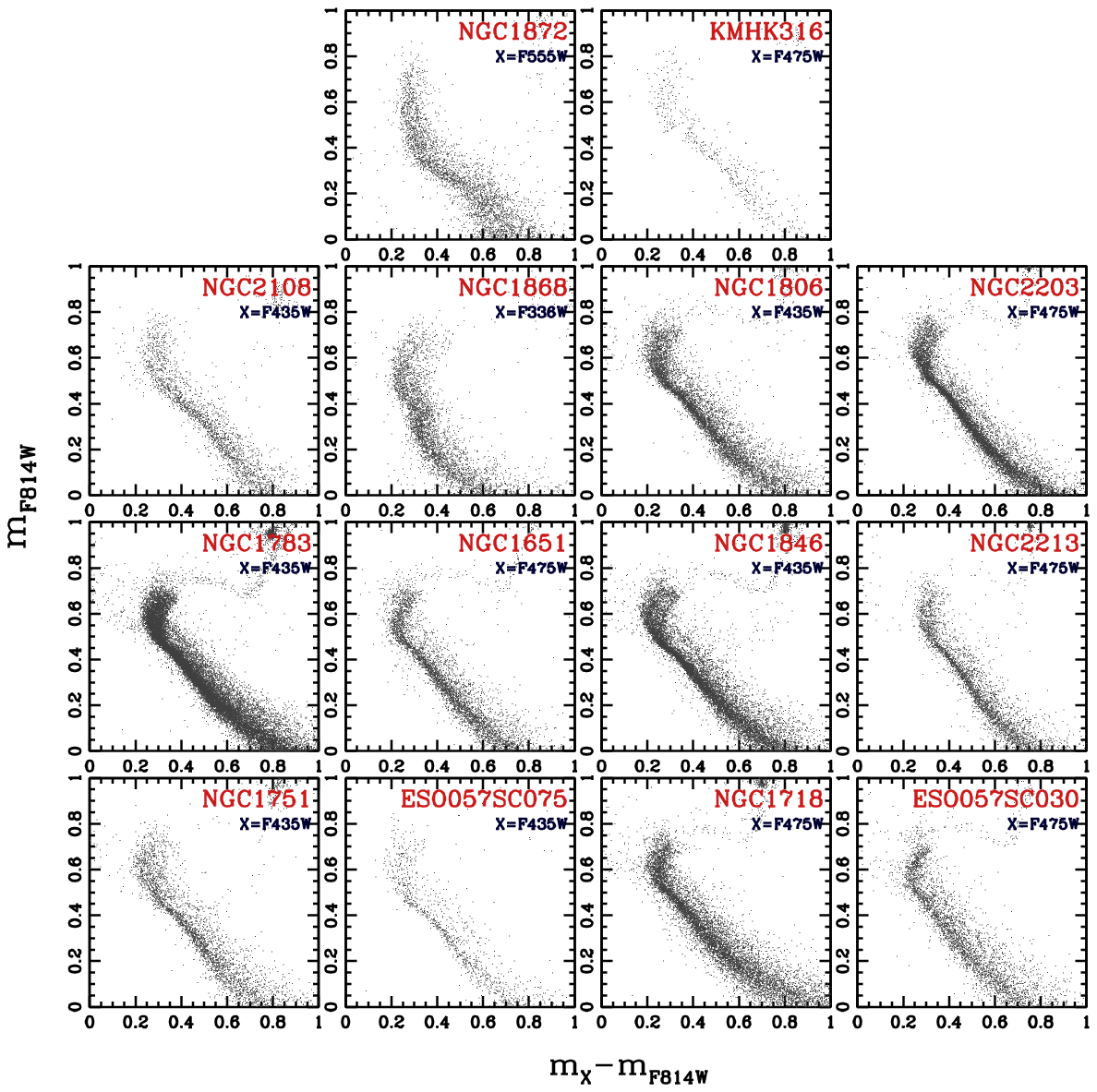}
\caption{Collection of CMDs for the investigated  star clusters. For each cluster, we plot the F814W magnitude against the X$-$F814W color, where the X filter is quoted in the corresponding panel.
 }
\label{fig:cmd}
\end{figure*}

\par This work investigates fourteen Magellanic Cloud star clusters younger than $\sim$2 Gyr that span an interval of mass (10$^4$ and $10^5$ solar masses) that is poorly explored in the context of binaries.
 To obtain robust results, our sample does not include clusters younger than $\sim$0.6 Gyr. 
 Indeed, these star clusters exhibit split MSs associated with stellar populations with different rotation rates \citep[][and references therein]{milone2022a}, which can alter our binary fraction analysis. 

\par The clusters are observed with the Wide Field Channel (WFC) of the ACS and the Ultraviolet and Visible channel (UVIS) of the Wide Field Camera 3 (WFC3) on board HST. The details of the observations are summarised in Table\,\ref{table:1}. 
We have corrected the effect of the poor Charge Transfer Efficiency (CTE) by using the method described in \cite{andersonbedin2010}. 
 
High-precision photometry and astrometry, essential for an accurate binary analysis, are derived using the state-of-the-art data reduction programs developed by Jay Anderson \citep[e.g.][]{andserson2000, anderson2008}, which are based on the  effective point spread function (ePSF) fitting. We used the FORTRAN software package, KS2, which is an upgraded version of the program kitchen$\_$sync \citep{anderson2008}, and it entails three different methods to measure the positions and magnitudes of stars;  
i) Method I measures the stars in each image, independently, by using the best available ePSF model. The results are then averaged to derive the best estimates of magnitude and position. This method provides the best photometry for relatively bright stars that define distinct peaks in a 5$\times$5 pixel raster. 
ii) Method II measures each star by means of aperture photometry, after subtracting all the neighboring stars. It works well for faint stars that do not have enough photons to be well-fitted by the ePSF. iii) Method III is similar to method II in terms of analysis and it works well in very crowded regions.
We refer to papers by \cite{sabbi2016a}, \cite{bellini2017a}, and \cite{milone2023} for details on KS2.

Since we are interested in stars with high-precision photometry, we  excluded all the sources that were poorly fitted by the PSF model and the stars with large root mean square values in positions. To select these stars, we used the computer programs and the methods provided by \cite{milone2012}.
The stellar coordinates are corrected for geometric distortion by using the solutions by \citet[][]{anderson2006} and \citet{bellini2011} the photometry has been calibrated to the Vega magnitude system as in \citet{milone2023} using the zero points provided by Space Telescope Science Institute (STScI) \footnote{\url{https://www.stsci.edu/hst/instrumentation/acs/data-analysis/zeropoints}}.
 
 The photometry was corrected for the effects of differential reddening by using the methods by \citet{milone2012} and \citet{legnardi2023a}. As an example, Figure\,\ref{fig:dr} compares the original CMD of NGC\,1751 (left)  with the CMD corrected for differential reddening (right).

Finally, we derived the ages, metallicities, distances, and reddening of the studied clusters by comparing the CMDs with isochrones, as in \citet{cordoni2023}.
We used the isochrone from the Dartmouth Stellar Evolution Database \footnote{\url{http://stellar.dartmouth.edu/models/isolf_new.html}} for clusters older than 1 Gyr and the 
MESA Isochrones and Stellar Tracks \footnote{\url{http://waps.cfa.harvard.edu/MIST/}} for clusters that are younger than 1 Gyr. The reddening coefficients for the different filters are provided by  Aaron Dotter (private communication).

We performed artificial-star (AS) tests to estimate the photometric errors, the level of completeness of our sample, and the fraction of blended sources that contaminate the CMD region populated by binaries. To do that, we used the method by \cite{anderson2008} and \cite{milone2009}.
 For each cluster, we generate a catalog of 100,000 ASs with instrumental magnitudes between $-4$ and $-13.7$ in the F814W bands. The colors of each AS are derived from the empirical fiducial line of MS stars. For the ASs, we adopted the same radial distribution and luminosity function as the real stars and reduced them using the same PSF model and procedure adopted for real stars. Moreover, we used the same criteria as for real stars to select the sample of relatively isolated ASs that are well-fitted by the PSF.
 Completeness is derived for each star as in \cite{milone2009} (see their section 2.2), by accounting for its magnitude and radial distance from the cluster center.

\section{Structural parameters of the star clusters}
\label{sec:parametri}

In this section, we discuss how we derived the density profile of the studied clusters and inferred the values of the core radius, density, and mass of each cluster. These quantities are crucial in properly characterising the populations of binaries in star clusters.

\subsection{Density profiles of the clusters}\label{sec:denprofit}

\begin{figure*}[ht]
    \centering
        \includegraphics[width=9cm]{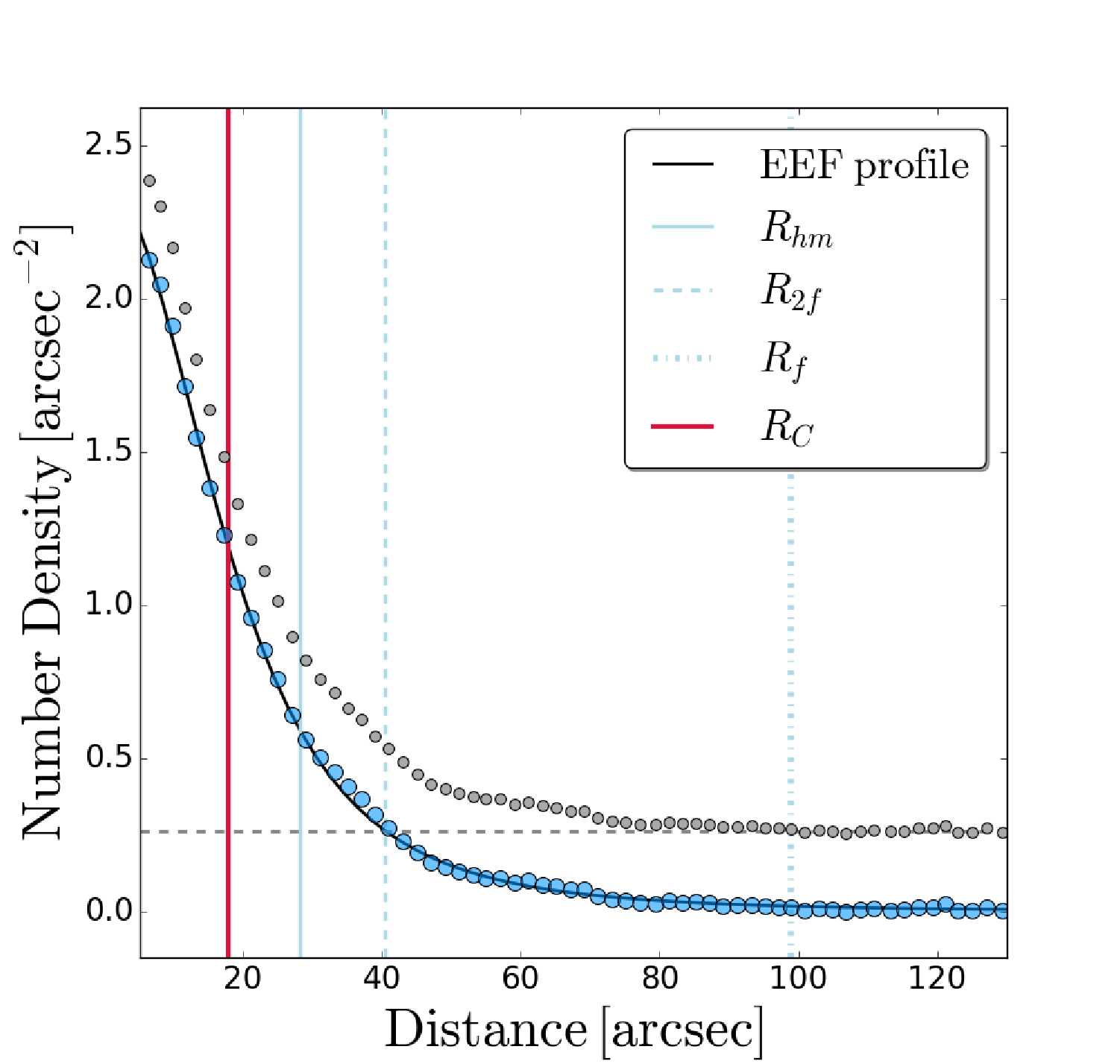}
        \includegraphics[width=9cm]{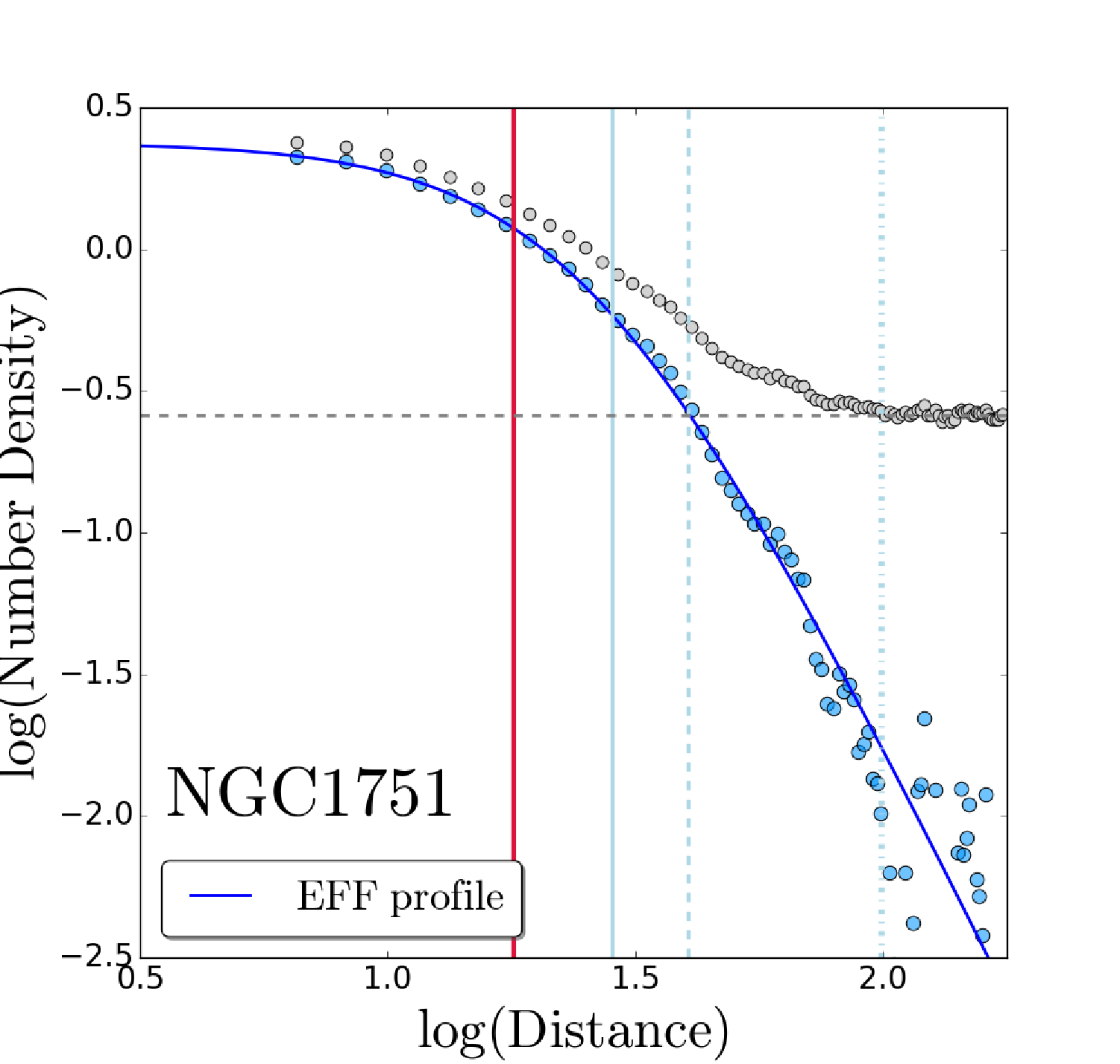}
        \caption{ Density profile. {\it Left Panel:} The grey points correspond to the observed density profile of the cluster NGC1751. The contribution from field contamination is denoted by the grey dotted horizontal line. The profile in blue is obtained after subtracting the background contamination from the actual density estimate, and it is fitted with the EEF profile. The {\it{right panel}} shows the corresponding quantities in a logarithmic scale. 
        The vertical lines indicate the core radius, the corresponding half-mass radius denoted by $R_{hm}$, and the distances indicated as $R_{f}$ and $R_{2f}$. The background level, bg, is marked by the horizontal line.}
        \label{fig:profilefit}
\end{figure*}

\begin{table*}
\caption{Parameters of the best fitting density profile.}
\label{table:densityprofile info.}
\centering
\resizebox{52em}{10em}{
\begin{tabular*}{1.2\textwidth}{@{\extracolsep{\fill}} c c c c c c c c c c c l}
\hline
\\
    &  \multicolumn{5}{c}{EEF profile fit} & \multicolumn{3}{c}{King's profile fit} &      &    \\
Cluster & $\mu_{0}$ &  $a$ & $\gamma$ & R$_{c}$ & $bg$ & K & R$_{c}$ & R$_{t}$  & R$_{hm}$ & $\rho_{R_{C}}$ & $\rho_{R_{hm}}$ \\
ID & [arcsec$^{-2}$] & [arcsec$^{-1}$] &  & [arcsec] & [arcsec$^{-2}$] & [arcsec$^{-2}$] & [arcsec] & [arcsec] &  [arcsec] & [arcsec$^{-2}$] & [arcsec$^{-2}$] \\
    \\
    \hline 
    \\[0.001em]
ESO057SC075   &   1.07 $\pm$   0.03  &  14.1 $\pm$  1.2  &  1.80 $\pm$  0.13  &  15 $\pm$  4   &    0.06     &   1.23 $\pm$  0.02  &  15.9 $\pm$  0.4  &  159 $\pm$  13  &      30    &   0.84    &  0.61    \\ 
ESO057SC030  &   4.49 $\pm$   0.06  &  15.5 $\pm$  0.9  &  1.54 $\pm$  0.08  &  19 $\pm$  2   &    0.45     &   5.29 $\pm$  0.08  &  19.6 $\pm$  0.3  &  142 $\pm$   6  &      31    &   3.54    &  2.84    \\
KMHK316      &   2.75 $\pm$   0.19  &   5.6 $\pm$  0.7  &  1.06 $\pm$  0.05  &   9 $\pm$  1   &    0.33     &   2.41 $\pm$  0.05  &   9.7 $\pm$  0.3  &  145 $\pm$  20  &      22    &   2.10    &  1.45    \\
NGC1651      &   3.31 $\pm$   0.01  &  19.3 $\pm$  0.3  &  2.10 $\pm$  0.03  &  19 $\pm$  2   &    0.14     &   4.23 $\pm$  0.04  &  20.6 $\pm$  0.2  &  146 $\pm$   4  &      32    &   2.63    &  2.02    \\
NGC1718      &   7.15 $\pm$   0.07  &  12.6 $\pm$  0.3  &  1.89 $\pm$  0.03  &  13 $\pm$  1   &    0.20     &   7.89 $\pm$  0.06  &  14.2 $\pm$  0.2  &  168 $\pm$   8  &      30    &   5.60    &  3.64    \\
NGC1751      &   2.83 $\pm$   0.05  &  14.1 $\pm$  0.7  &  1.44 $\pm$  0.04  &  18 $\pm$  1   &    0.26     &   3.40 $\pm$  0.04  &  19.8 $\pm$  0.2  &  128 $\pm$   3  &      28    &   2.23    &  1.81    \\
NGC1783      &   4.82 $\pm$   0.03  &  33.3 $\pm$  0.9  &  2.34 $\pm$  0.07  &  30 $\pm$  8   &    0.26     &   7.27 $\pm$  0.13  &  34.7 $\pm$  0.4  &  161 $\pm$   4  &      41    &   3.84    &  3.31    \\
NGC1806      &   3.92 $\pm$   0.02  &  27.3 $\pm$  0.9  &  2.30 $\pm$  0.09  &  25 $\pm$  8   &    0.36     &   5.43 $\pm$  0.07  &  26.2 $\pm$  0.3  &  138 $\pm$   4  &      33    &   3.12    &  2.74    \\
NGC1846      &   1.89 $\pm$   0.02  &  22.0 $\pm$  0.9  &  1.73 $\pm$  0.06  &  24 $\pm$  3   &    0.10     &   2.29 $\pm$  0.05  &  26.9 $\pm$  0.5  &  208 $\pm$  13  &      44    &   1.48    &  1.12    \\
NGC1868      &   6.17 $\pm$   0.06  &  11.7 $\pm$  0.3  &  2.39 $\pm$  0.04  &  10 $\pm$  2   &    0.07     &   7.48 $\pm$  0.04  &  11.1 $\pm$  0.1  &  115 $\pm$   4  &      21    &   4.92    &  3.39    \\
NGC1872      &   8.48 $\pm$   0.21  &   9.4 $\pm$  0.5  &  1.51 $\pm$  0.06  &  12 $\pm$  1   &    0.82     &  10.09 $\pm$  0.20  &  12.5 $\pm$  0.3  &   83 $\pm$   4  &      16    &   6.69    &  5.76    \\
NGC2108      &   2.81 $\pm$   0.04  &  13.3 $\pm$  0.5  &  1.43 $\pm$  0.04  &  17 $\pm$  1   &    0.29     &   3.15 $\pm$  0.03  &  17.5 $\pm$  0.2  &  143 $\pm$   5  &      29    &   2.22    &  1.72    \\
NGC2203      &   2.42 $\pm$   0.01  &  30.8 $\pm$  0.8  &  3.03 $\pm$  0.09  &  23 $\pm$ 17   &    0.02     &   3.60 $\pm$  0.06  &  27.3 $\pm$  0.4  &  148 $\pm$   5  &      36    &   1.93    &  1.57    \\
NGC2213      &   3.59 $\pm$   0.07  &  12.2 $\pm$  0.6  &  2.24 $\pm$  0.08  &  11 $\pm$  3   &    0.06     &   4.20 $\pm$  0.05  &  11.8 $\pm$  0.2  &  145 $\pm$  10  &      24    &   2.84    &  1.90    \\
\\[0.001em] \hline
    \end{tabular*}%
}
\end{table*}

To estimate the density profile of each star cluster we used the procedure illustrated in Figure\,\ref{fig:profilefit} for the cluster NGC\,1751.
We first derived the number of stars with mass greater than $0.9 M_{\odot}$ in different annuli in the HST FoV and then corrected those numbers for completeness. The obtained numbers are normalised by the areas of the annuli to get the number density profile of the cluster. 

We derived the parameters of the EEF profile \citep[][]{elson1987} that provide the best-fit with the observed density profile. 
Specifically, we adopted the relation,
\begin{equation}
\label{eq:eef}
    \mu(r)=\mu_{0}(1 + \frac{r^2}{a^2})^{-\frac{\gamma}{2}} +bg
\end{equation}

\par where $\mu_{0}$ is the central density, $a$ is the scaling factor, $\gamma$ is the power law factor, and $bg$ is the constant that accounts for the contamination from field stars. The core radius, $R_{c}$ is derived by the relation, 
\begin{equation}
\label{eq:eefrc}
    R_{c}=a({2^{\frac{2}{\gamma}} - 1})^\frac{1}{2} 
\end{equation}

The left panel of Figure \ref{fig:profilefit} shows the EEF density profile that provides the best fit with the observations of NGC\,1751. The right panel shows the corresponding plot in the logarithmic scale, i.e., EFF Profile. The grey and blue points denote the number density distribution with and without the contamination from field stars. Blue points are obtained by subtracting $bg$ from the grey points, where the black horizontal line denotes the level of $bg$. The corresponding radius is $R_{f}$. The radius at which the number density becomes two times $bg$ is denoted as $R_{2f}$.

For comprehensiveness, we derived the King profile \citep[][]{king1962} parameters that provide the best-fit with the observed density profile. The results are listed in Table \ref{table:densityprofile info.}. 

\subsection{Mass functions and Masses of the clusters}

\begin{table*}
\caption{Parameters of the Mass function along with the total mass of the cluster.}
\label{table:cluster mass}
\centering
    \begin{tabular*}{0.6\textwidth}{@{\extracolsep{\fill}}  c c c c c }
    \hline
    \hline
   Cluster & $\alpha$ & $\bar{M}$ & $M_{Cluster}$ & $t_{rh}$ \\
ID &  &  [M$\odot$] & [M$\odot$] & [Gyrs] \\
    \\
    \hline
ESO057SC075  &  -1.56 $\pm$  0.11  &    0.77    &  9.86  $\times$   $10^{3}$  &  0.27   \\ 
ESO057SC030  &  -2.15 $\pm$  0.04  &    0.95    &  8.77  $\times$   $10^{4}$  &  0.51   \\ 
KMHK316      &  -2.09 $\pm$  0.06  &    1.17    &  1.42  $\times$   $10^{4}$  &  0.12   \\ 
NGC1651      &  -1.53 $\pm$  0.07  &    1.00    &  3.72  $\times$   $10^{4}$  &  0.46   \\
NGC1718      &  -1.95 $\pm$  0.04  &    0.99    &  7.64  $\times$   $10^{4}$  &  0.50   \\ 
NGC1751      &  -1.74 $\pm$  0.04  &    1.04    &  3.05  $\times$   $10^{4}$  &  0.30   \\ 
NGC1783      &  -1.48 $\pm$  0.05  &    0.79    &  8.15  $\times$   $10^{4}$  &  1.15   \\ 
NGC1806      &  -1.38 $\pm$  0.12  &    1.15    &  4.67  $\times$   $10^{4}$  &  0.50   \\ 
NGC1846      &  -1.29 $\pm$  0.07  &    0.81    &  4.12  $\times$   $10^{4}$  &  0.68   \\ 
NGC1868      &  -1.28 $\pm$  0.13  &    1.23    &  1.61  $\times$   $10^{4}$  &  0.14   \\ 
NGC1872      &  -2.81 $\pm$  0.05  &    1.13    &  8.24  $\times$   $10^{4}$  &  0.14   \\ 
NGC2108      &  -2.30 $\pm$  0.06  &    1.11    &  4.34  $\times$   $10^{4}$  &  0.31   \\ 
NGC2203      &  -1.63 $\pm$  0.04  &    0.79    &  3.35  $\times$   $10^{4}$  &  0.58   \\ 
NGC2213      &  -1.96 $\pm$  0.06  &    1.13    &  2.44  $\times$   $10^{4}$  &  0.21   \\
\hline
\hline
    \end{tabular*}
\end{table*}

\par To derive the mass functions of each cluster, we used the procedure by \citet{cordoni2023}. We derived the number of MS stars within the radial distance of $R_{2f}$, divided them into intervals of equal mass, and normalized this quantity to the mass bin.
The star counts are corrected for completeness, and only stars with completeness values larger than 0.5 are taken into account.

We fitted the observed mass function with the relation by \citet{trenti2010}.
\begin{equation}
\label{eq:mass}
    \xi_{m}=\xi_{0} \cdot m^{-\alpha}
\end{equation}

To estimate the total mass of the cluster, $M_{Cluster}$, we integrated the mass function over the entire mass interval provided by the best-fit isochrone. Moreover, we accounted for the cluster stars with a radial distance larger than $R_{2f}$ by using the cluster EEF profile and assuming a homogeneous radial distribution for stars with different masses.

Finally, we estimated the half-mass relaxation time by using the following equation,

\begin{equation}
    \label{eq:trh}
    t_{rh}= \frac{ 0.138 \cdot M_{Cluster}^{1/2} \cdot {R_{hm}}^{3/2} }{G \cdot \bar{M} \cdot ln( 0.11 \cdot M_{Cluster} / \bar{M} ) }
\end{equation}
where $G$ is the universal gravitational constant, $\bar{M}$ is the average mass of a star, and $R_{hm}$ is the half-mass radius \citep{spitzer1987}. The results are provided in Table\,\ref{table:cluster mass}. 

\section{The binary fraction}\label{sec:method}

\begin{figure}[ht]
\centering
\includegraphics[width=7.5cm]{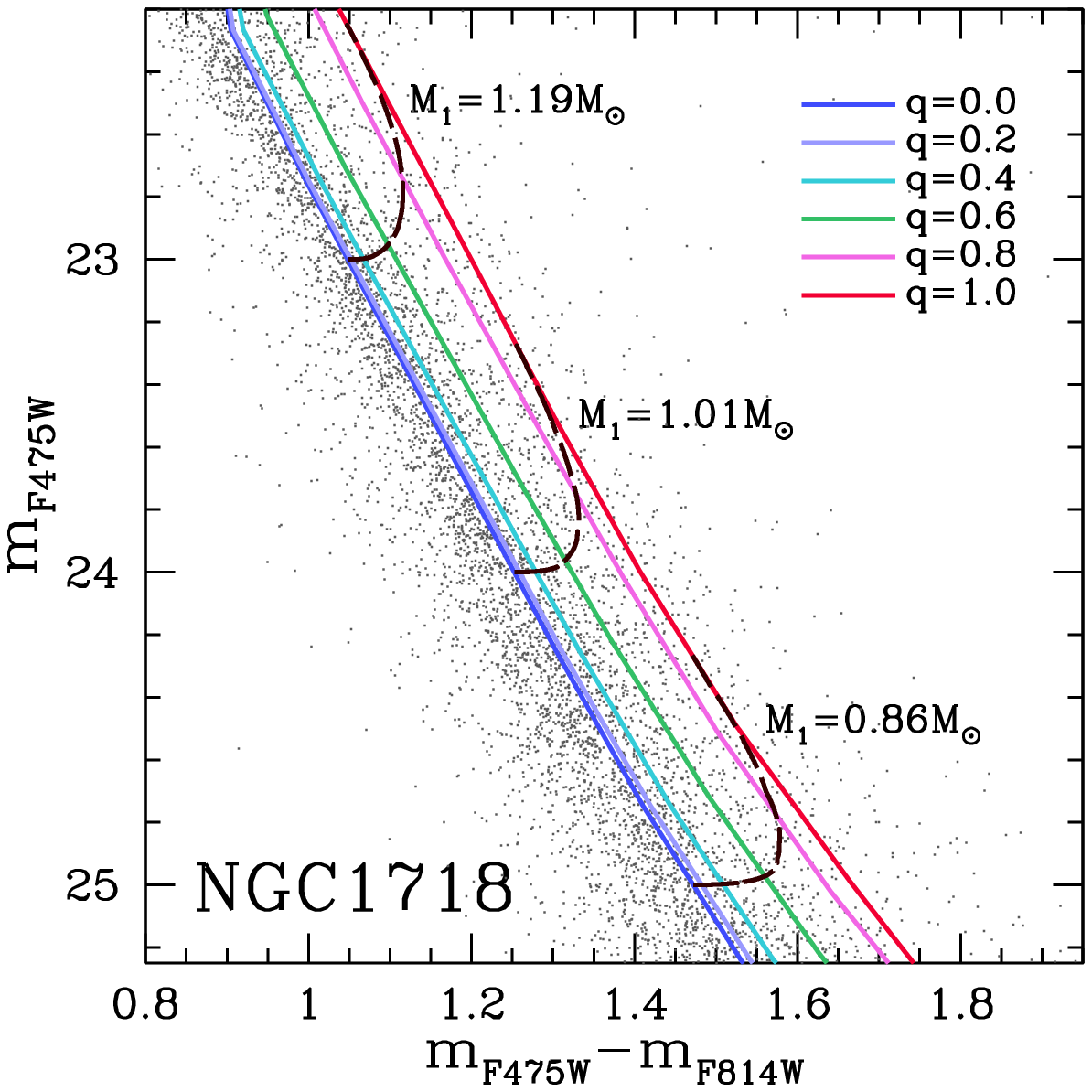}
\caption{
Loci of binaries. CMD of NGC\,1718 zoomed around the upper MS. The colored continuous lines are the fiducial lines of binary systems with different mass ratios, as indicated in the inset. The dashed lines represent the locus of binaries with primary masses  of 1.19, 1.01, and 0.86 solar masses and mass ratios between 0 and 1.}
\label{fig:q1curves}
\end{figure}

\begin{table*}
\caption{Parameters for the best fitting isochrone.}
\label{table:details}
\centering
    \begin{tabular*}{0.7\textwidth}{@{\extracolsep{\fill}} c c c c c c c}
    \hline
    \hline
   Cluster  & Age & [Fe/H] & (m$-$M)$_{0}$ & E(B-V) & RA & DEC \\
ID & [Gyr] & [dex] & [mag] & [mag] & h m s & d m s \\
    \\
    \hline
ESO057SC075      & 1.90 & -0.40 & 18.49 & 0.06 & 06 13 27.26 & -70 41 45.0     \\
ESO057SC030     & 2.10 & -0.50 & 18.40 & 0.17 & 05 42 17.65 & -71 35 28.2     \\ 
KMHK316         & 0.90 & -0.30 & 18.35 & 0.12 & 04 56 37.46 & -68 09 55.8     \\ 
NGC1651         & 1.70 & -0.40 & 18.70 & 0.14 & 04 37 32.23 & -70 35 10.8     \\ 
NGC1718         & 1.90 & -0.50 & 18.53 & 0.23 & 04 52 25.89 & -67 03 06.6     \\ 
NGC1751         & 1.75 & -0.50 & 18.55 & 0.15 & 04 54 11.99 & -69 48 27.1     \\ 
NGC1783         & 1.60 & -0.39 & 18.75 & 0.07 & 04 59 08.97 & -65 59 13.8     \\ 
NGC1806         & 1.50 & -0.40 & 18.78 & 0.10 & 05 02 11.72 & -67 59 08.0     \\ 
NGC1846         & 1.70 & -0.50 & 18.25 & 0.12 & 05 07 34.15 & -67 27 36.7     \\ 
NGC1868         & 1.45 & -0.40 & 18.45 & 0.06 & 05 14 35.91 & -63 57 15.1     \\ 
NGC1872         & 0.60 & -0.40 & 18.31 & 0.18 & 05 13 11.29 & -69 18 44.9     \\ 
NGC2108         & 1.00 & -0.30 & 18.40 & 0.14 & 05 43 56.54 & -69 10 52.9     \\ 
NGC2203         & 1.55 & -0.30 & 18.55 & 0.10 & 06 04 42.62 & -75 26 16.1     \\ 
NGC2213         & 1.70 & -0.40 & 18.50 & 0.10 & 06 10 42.13 & -71 31 45.9     \\  
\hline
\hline
\end{tabular*}
\end{table*}

\begin{figure*}[ht]
\begin{center}
    \includegraphics[width=16cm]{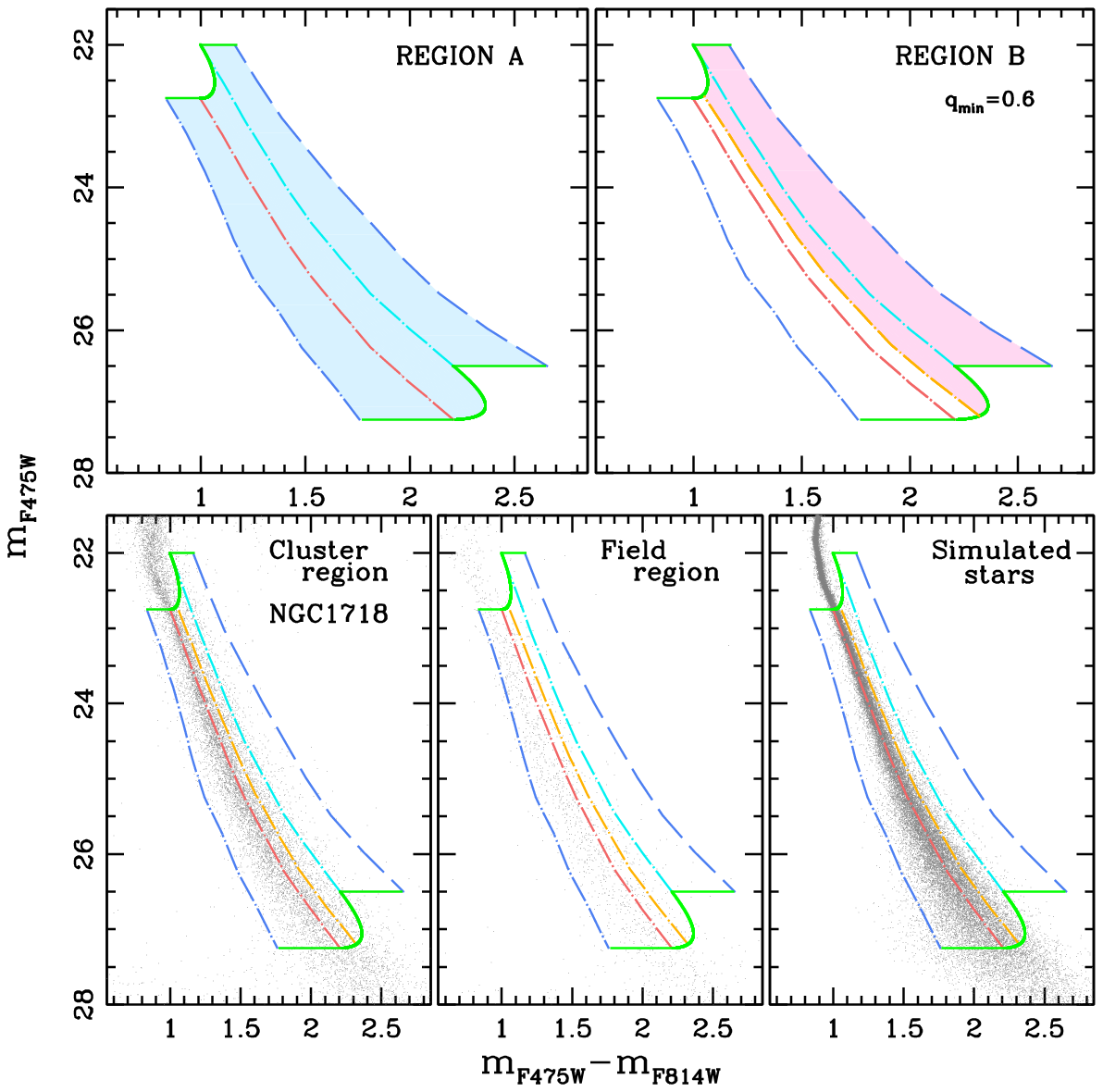}
\end{center}
\caption{Method of analysis. The figure summarises the method that was used to analyse MS-MS binaries in NGC\,1718. 
\textit{Top}. The {\it{top-left panel}} highlights the Region A of the $m_{\rm F475W}$ vs.\,$m_{\rm F475W}-m_{\rm F814W}$ CMD, which is shaded with light-blue colour. Region B, which is a subsection of Region A, is shaded with a pink color in the {\it{top-right panel}}. CMDs of stars in the cluster region ({\it{bottom-left panel}}), field region ({\it{bottom-centre panel}}), and simulated CMD ({\it{bottom-right panel}}). 
The MS fiducial line and the fiducial of equal-mass binaries are represented with red and cyan dot-dashed lines, respectively, in all panels. The blue-dot--dashed and blue-dashed lines mark the left and right color boundaries of region A, displaced by three times the colour error from the fiducial and equal mass binary lines respectively. Green curves represent the mass limit of analysis, i.e., binaries with certain primary mass and the mass ratio with secondary mass ranging between 0 and 1. Orange lines represent the unequal mass binary line, i.e., binaries with a mass ratio, $q=0.6$. See the text for details.
}
\label{fig:method}
\end{figure*}

\par Star clusters are dense systems. Surviving in such environments requires the stars in a binary to be in close proximity. Even with the high-resolution power of HST, it is hard to resolve the stars in such a system. Hence, the binaries are perceived as single stars of enhanced magnitude, where the observed flux will be the combination of the fluxes of the companion stars.
\par Suppose we observe a binary system where the magnitudes and fluxes of the companion stars are $m_{1}$ and $m_{2}$, and $F_{1}$ and $F_{2}$, respectively. Then, the magnitude of the binary system is,

\begin{equation}
\label{eq:1}
    m_{bin}=m_{1}-2.5log(1+\frac{F_{2}}{F_{1}})
\end{equation}

where subscripts 1 and 2 denote the primary and secondary stars, respectively, with the primary star being the more massive component of the binary system.  

In a simple stellar population, the fluxes of MS stars depend on stellar mass and follow a given mass-luminosity relation. Hence, the luminosity of a binary system depends on the mass ratio between the companion stars,
\begin{equation}
\label{eq:2}
    q=\frac{M_{2}}{M_{1}}
\end{equation}

\par where, $0 < q \leq 1$.

As illustrated in Figure\,\ref{fig:q1curves}, MS-MS binaries populate the CMD region on the red and bright side of the MSFL, with equal-mass binaries defining a fiducial line that is parallel to the MSFL but is shifted by $0.7526$ mag in brightness.
The fiducial lines are derived from the observed MS stars of the cluster NGC\,1718, which is taken as a test case for demonstration. Specifically, we selected the well-measured candidate MS stars and divided this sample into intervals of 0.5 mag in the F475W band. The MSFL is derived by linearly interpolating the median colors and magnitudes of the stars in each bin. The standard deviation of the colors of stars in each bin, $\sigma$, is considered a proxy for the average color error of the stars in that bin.

For a fixed mass of the primary star, the binaries with $q<1$ distribute on a curved line between the MSFL and the equal-mass binary fiducial line. Due to observational errors, it is not possible to disentangle binaries with small mass ratios from single stars in the observed CMD. 
The colors and magnitudes of the binaries with different mass ratios are determined with the mass–luminosity relation provided by the best-fitting isochrones provided by \cite{dotter2008a, dotter2016a}. The latter is derived by comparing the CMD of each cluster with a grid of solar-scaled isochrones  that account for different ages, metallicities ([Fe/H]), distance modulus ((m$-$M)$_{0}$), and reddening \citep[E(B$-$V), see][for details]{milone2009}. The distance to the cluster is calculated from (m$-$M)$_{0}$. The values of age, [Fe/H], (m$-$M)$_{0}$, and E(B$-$V) that provide the best match with the observed data are listed in Table \ref{table:details}. \footnote{The ages obtained in our work are in agreement, at 1-$\sigma$ level, with those by \cite{sun2018}, who compared the observed CMDs with Padova group’s PARSEC 1.2S isochrones \citep{bressan2012}. The average age difference is 190 Myr, which is comparable with the age error of $\sim 200$ Myr by Sun and collaborators. The distance modulus, (m$-$M)$_{0}$ derived in our work differ, on average, by 0.18 mag with those by \cite{sun2018}.} 

\par The method to derive the fraction of binaries is illustrated in Figure\,\ref{fig:method} for NGC\,1718.
We divided the CMD into two regions. Region \,A includes all the single stars in the studied luminosity interval and the binary systems with a primary star in the same magnitude range. It is limited to the left by the MSFL blue-shifted by three times the color observational error ($\sigma$). The right boundary corresponds to the fiducial line of equal-mass binaries red-shifted by three times the color error.
The upper and lower boundaries are the sequences of binaries where the mass ratio ranges from zero to one and the primary stars have  magnitudes  $m_{\rm F475W}=m_{\rm F475W}^{\rm bright}$ and $m_{\rm F475W}^{\rm faint}$, respectively. Here, $m_{\rm F475W}^{\rm bright}$ and  $m_{\rm F475W}^{\rm faint}$, corresponding to masses $M_{u}$ and $M_{l}$, are the bright and faint limit of the studied luminosity interval. $m_{\rm F475W}^{\rm bright}$ is selected such that region A does not intervene with the turnoff region and the selected $m_{\rm F475W}^{\rm faint}$ assures the completeness of the magnitude range of analysis remains above 0.5. Region B is the portion of Region A that is predominantly populated by binaries with $q$ $\geq$ $q_{min}$. The value of $q_{\rm min}$ is chosen with the criterion that the fiducial line of binaries with $q=q_{\rm min}$ is redder than the MSFL shifted by $\sigma$ to the red.

\par  Moreover, to derive the fraction of binaries, we defined two regions in the HST FoV. 
1) The cluster-region that extends from the dense cluster centre to $R_{2f}$. 
2) Starting from $R_{2f}$, the field-region stretches to the outermost parts of the HST FoV. Given the small FoV of HST, we can assume that the distributions of field stars in the cluster-region and in the field-region are nearly the same. The field stars appearing in cluster-region may contribute to the stars present in regions A and B of the cluster-region CMD. Their amount is estimated by constructing a CMD using the stars appearing in the field-region. 
As an example, the bottom-left and bottom-middle panels of Figure\,\ref{fig:method} show the CMDs of stars in the cluster-region and the field-region of NGC\,1718, respectively.

\par The binary fraction is derived using the Equation\,1 of \cite{milone2012}.
\begin{equation} 
\label{eq:3}
F_{\rm bin}^{q \geq q_{\rm min}}=\frac{N_{\rm cluster}^{\rm B}-N_{\rm field}^{\rm B}}{N_{\rm cluster}^{\rm A}-N_{field}^{\rm A}}-\frac{N_{\rm art}^{\rm B}}{N_{\rm art}^{\rm A} }
\end{equation}

Here, $N_{\rm cluster}^{\rm A}$ and $N_{\rm cluster}^{\rm B}$ are the number of stars in the cluster-region, corrected for completeness, in the regions A and B of the CMD, respectively. $N_{\rm field}^{\rm A}$ and $N_{\rm field}^{\rm B}$ are the corresponding number of stars in the field-region normalized by the ratio between the respective areas of the cluster and field region. 
$N_{\rm art}^{\rm A}$ and $N_{\rm art}^{\rm B}$ are the numbers of artificial stars that populate the regions A and B of the CMD. 
The artificial star binary fraction is denoted by the term $\frac{N_{\rm art}^{\rm B}}{N_{\rm art}^{\rm A}}$ of the Equation\,\ref{eq:3}. This term accounts for the fraction of single stars that populate region B of the CMD due to large photometric errors or chance superposition of their images with other single stars.

\section{Results and Discussions} \label{sec:floats}

\begin{table*}
\caption{Fraction of binaries for all studied clusters. We provide the radius of the studied region and the stellar-mass interval.
\label{table:fraction}}
\centering
    \begin{tabular*}{0.9\textwidth}{@{\extracolsep{\fill}}  c c c c c c c }
    \hline
    \hline
   Cluster & F$_{bin, R_{hm}}^{q>0.6}$  & F$_{bin, R_{hm}}^{q>0.7}$ & F$_{bin, core}^{q>0.6}$ & F$_{bin, core}^{q>0.7}$ & M$_{u}$ & M$_{l}$ \\
ID &  &  &  &  & [$M\odot$] & [$M\odot$]  \\
    \\
    \hline
ESO057SC075 &        --------            &    0.150   $\pm$   0.013  &       --------                &   0.220   $\pm$  0.029     &  1.25   &  0.72   \\
ESO057SC030 &        --------            &    0.123   $\pm$   0.010  &       --------                &   0.103   $\pm$  0.017     &  1.16   &  0.65   \\
KMHK316     &        --------            &    0.105   $\pm$   0.023  &       --------                &   0.108   $\pm$  0.039     &  1.46   &  0.89   \\
NGC1651     &        --------            &    0.110   $\pm$   0.010  &       --------                &   0.117   $\pm$  0.016     &  1.29   &  0.81   \\
NGC1718     &    0.122   $\pm$   0.010   &    0.108   $\pm$   0.008  &     0.125     $\pm$  0.021    &   0.128   $\pm$  0.017     &  1.24   &  0.73   \\
NGC1751     &    0.221   $\pm$   0.022   &    0.175   $\pm$   0.017  &     0.213     $\pm$  0.031    &   0.171   $\pm$  0.024     &  1.21   &  0.74   \\
NGC1783     &    0.113   $\pm$   0.010   &    0.094   $\pm$   0.007  &     0.118     $\pm$  0.013    &   0.103   $\pm$  0.009     &  1.33   &  0.80   \\
NGC1806     &        --------            &    0.115   $\pm$   0.010  &       --------                &   0.112   $\pm$  0.013     &  1.39   &  0.86   \\
NGC1846     &    0.105   $\pm$   0.011   &    0.074   $\pm$   0.009  &     0.113     $\pm$  0.017    &   0.080   $\pm$  0.013     &  0.98   &  0.66   \\
NGC1868     &    0.112   $\pm$   0.018   &    0.093   $\pm$   0.014  &     0.108     $\pm$  0.042    &   0.107   $\pm$  0.033     &  1.32   &  0.87   \\
NGC1872     &         --------           &    0.111   $\pm$   0.020  &       --------                &   0.135   $\pm$  0.030     &  1.56   &  1.06   \\
NGC2108     &         --------           &    0.241   $\pm$   0.018  &       --------                &   0.237   $\pm$  0.026     &  1.44   &  0.76   \\
NGC2203     &    0.085   $\pm$   0.009   &    0.075   $\pm$   0.007  &     0.080     $\pm$  0.013    &   0.074   $\pm$  0.010     &  1.33   &  0.65   \\
NGC2213     &    0.132   $\pm$   0.016   &    0.119   $\pm$   0.012  &     0.146     $\pm$  0.028    &   0.133   $\pm$  0.023     &  1.27   &  0.66   \\
\hline
\hline
    \end{tabular*}
\end{table*}

\par For each cluster, we derived the fraction of binaries with a mass ratio, $q \geq 0.7$ by following the procedure described in Section\,\ref{sec:method}. Moreover, in the clusters where the binaries are better distinguishable from single stars, namely NGC\,1718, NGC\,1751, NGC\,1783, NGC\,1846, NGC\,1868, NGC\,2203, and NGC\,2213, we also studied binaries with $q \geq 0.6$.

\par The results are listed in Table\,\ref{table:fraction}, where we provide the fraction of binaries within the radial distance of $R_{hm}$ and the analyzed stellar-mass interval, given by the difference between the corresponding $M_{u}$ and $M_{l}$. The binary fraction ranges between $\sim$ 0.07 in NGC\,1846, to $\sim$ 0.24, in NGC\,2108. Table\,\ref{table:fraction} also provides the fraction of binaries within the core of each cluster \footnote{ The results of 12 out of the 14 analysed clusters are derived using the isochrones from Dartmouth Stellar Evolution databases, which are available for ages greater than 1 Gyr. In the cases of KMHK\,316 and NGC\,1872, which are younger than 1 Gyr, we used the MIST isochrones. To investigate possible systematic errors in the binary fraction due to the difference in the used isochrone, we calculated the fraction of binaries in three clusters, of different ages, by using both types of isochrones. In the $\sim 1$ Gyr old cluster NGC\,2108, the fraction of binaries within $R_{C}$ and $R_{hm}$ that we derived using the Dartmouth and the MIST isochrones differ by less than 1\%, whereas in NGC\,1868 (age of $\sim$1.5 Gyr) and ESO\,057SC030 (age of $\sim$2 Gyr) the fraction of binaries derived from the two sets of isochrones vary by $\sim$0.6\% and $\sim$0.3\%, respectively. Since such differences are much smaller than the corresponding observational errors, we conclude the results are not significantly affected by the adopted isochrones. }.

\begin{figure}[ht]
\centering
\includegraphics[width=8cm]{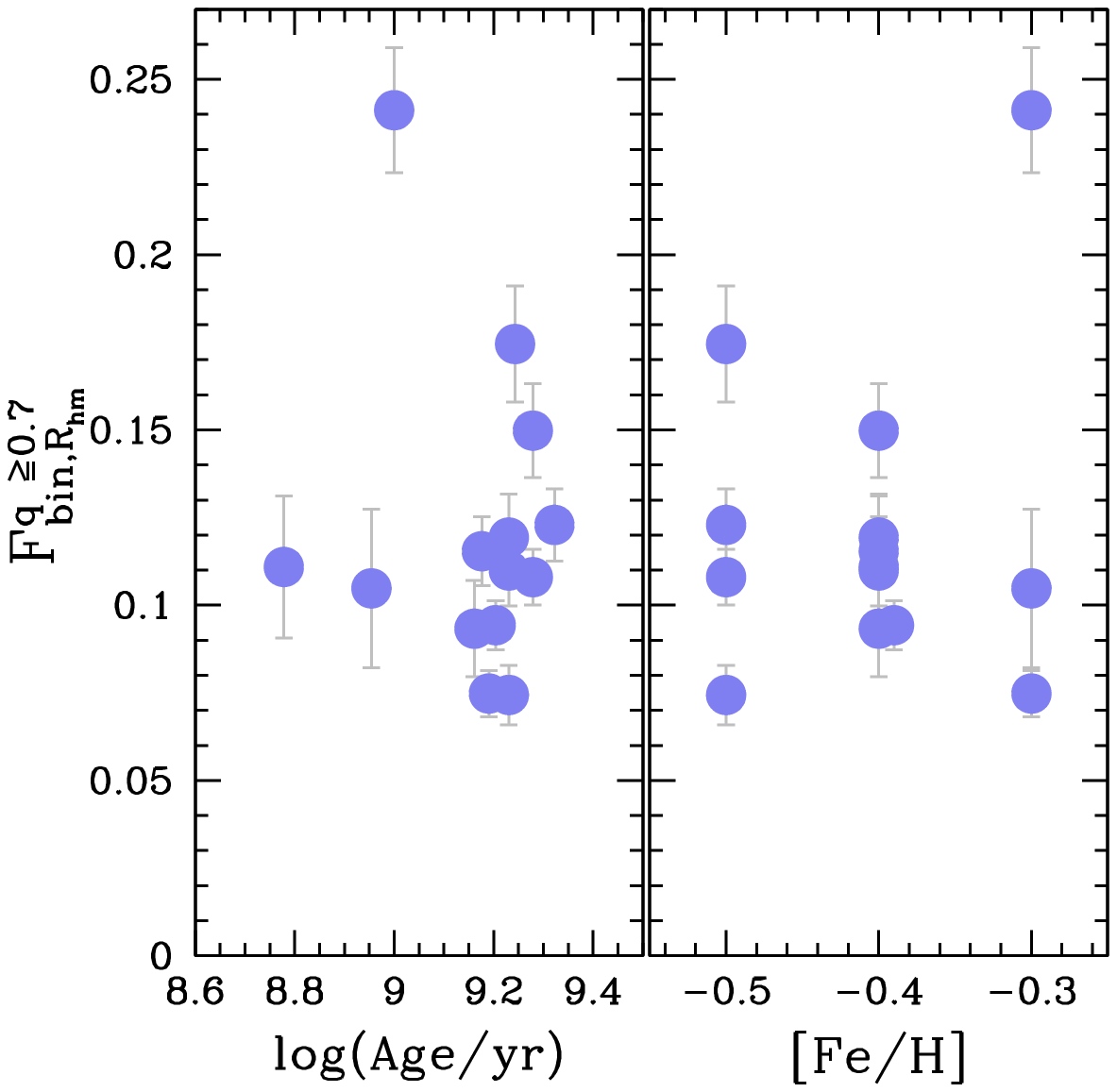}
\caption{Binary fraction and cluster parameters. Fraction of binaries with q$\geq$0.7 and within the $R_{hm}$ of the cluster is plotted against the age ({\it{left panel}}) and iron abundance ({\it{right panel}}) of the host cluster.} 
\label{fig:agemet}
\end{figure}

As shown in Figure\,\ref{fig:agemet}, the binary fraction exhibits  no significant correlations with cluster age and metallicity. This conclusion is supported by Spearman's rank correlation coefficients of 0.2 and -0.2, respectively.
 
\subsection{Binary fraction and the mass of the primary star}
\begin{figure*}[ht]
\includegraphics[width=18cm]{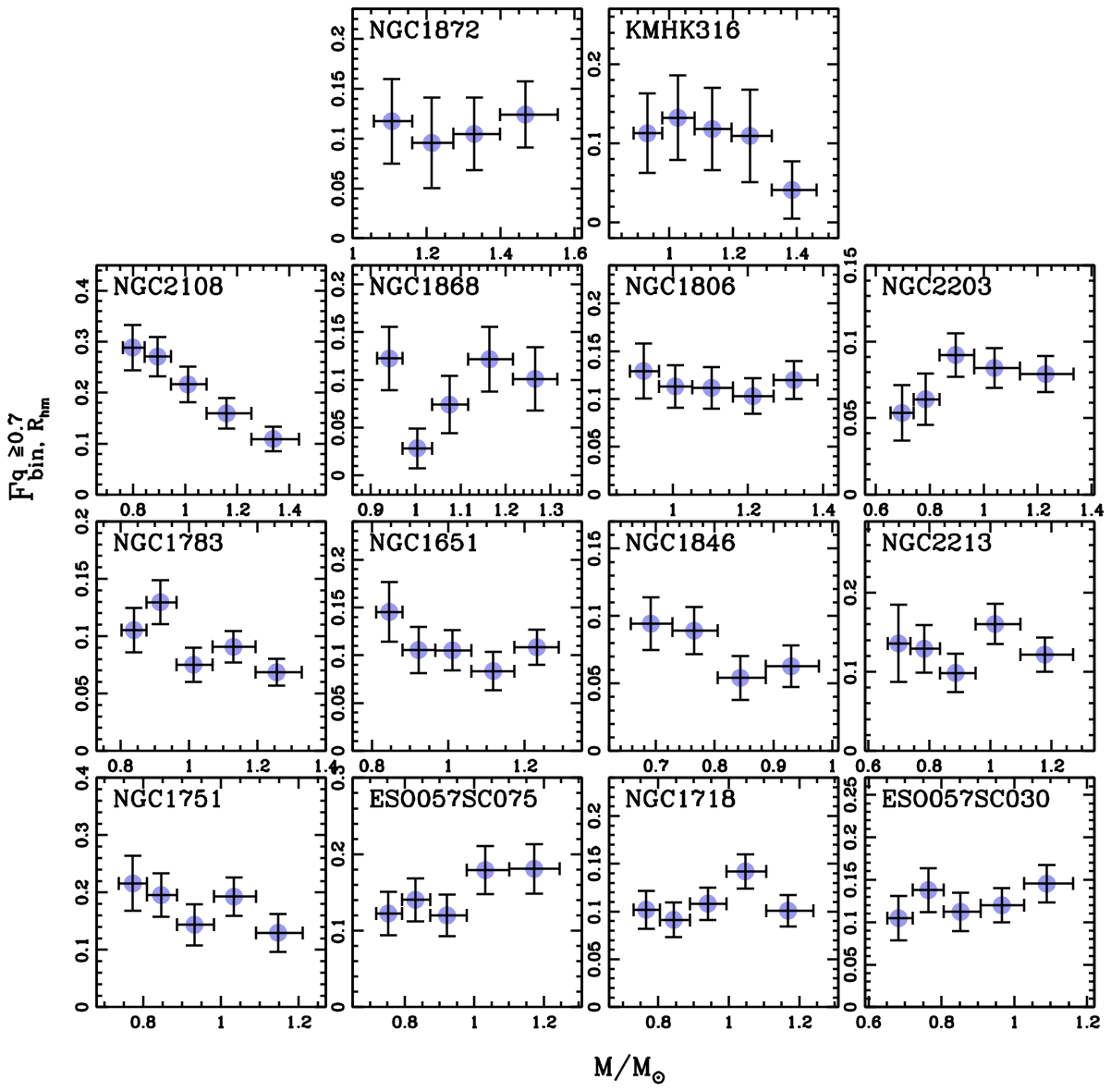}
\caption{Binary fraction and mass of the primary star. The fraction of binaries with $q \geq 0.7$ is plotted against the mass of the primary star. 
 }
\label{fig:mass}
\end{figure*} 

\begin{table*}
\caption{ $\chi^{2}$ and P-value derived for the relations between the  binary  fraction and the stellar mass, mass ratio, and radial distribution.}
\label{table:chi}
\centering    
\begin{tabular*}{0.7\textwidth}{@{\extracolsep{\fill}} c c c c c c c }
\hline
\hline
Cluster &  \multicolumn{2}{c}{Mass distribution} & \multicolumn{2}{c}{q distribution} & \multicolumn{2}{c}{Radial distribution}\\
ID & $\chi^{2}$ & P value & $\chi^{2}$ & P value & $\chi^{2}$ & P value \\ 
\\
\hline
ESO057SC075   &    7.1  $\times$ $10^{-4}$   &        0.40   &   11.6  $\times$ $10^{-3}$   &      0.02    &   11.0  $\times$ $10^{-4}$  &      0.05  \\
ESO057SC030   &    2.3  $\times$ $10^{-4}$   &        0.70   &   10.0  $\times$ $10^{-3}$   &      0.00    &    3.0  $\times$ $10^{-4}$  &      0.14  \\
KMHK316       &   10.1  $\times$ $10^{-4}$   &        0.72   &    4.1  $\times$ $10^{-3}$   &      0.30    &    2.9  $\times$ $10^{-4}$  &      0.91  \\
NGC1651       &    4.0  $\times$ $10^{-4}$   &        0.45   &    0.3  $\times$ $10^{-3}$   &      0.60    &    1.3  $\times$ $10^{-4}$  &      0.45  \\
NGC1718       &    3.0  $\times$ $10^{-4}$   &        0.32   &   15.0  $\times$ $10^{-3}$   &      0.00    &    2.0  $\times$ $10^{-4}$  &      0.07  \\
NGC1751       &   10.9  $\times$ $10^{-4}$   &        0.45   &   12.2  $\times$ $10^{-3}$   &      0.00    &    0.6  $\times$ $10^{-4}$  &      0.93  \\
NGC1783       &    4.8  $\times$ $10^{-4}$   &        0.05   &    6.1  $\times$ $10^{-3}$   &      0.00    &    1.2  $\times$ $10^{-4}$  &      0.10  \\
NGC1806       &    0.8  $\times$ $10^{-4}$   &        0.95   &    4.2  $\times$ $10^{-3}$   &      0.00    &    2.5  $\times$ $10^{-4}$  &      0.21  \\
NGC1846       &    3.0  $\times$ $10^{-4}$   &        0.29   &    0.4  $\times$ $10^{-3}$   &      0.44    &    1.9  $\times$ $10^{-4}$  &      0.10  \\
NGC1868       &   12.4  $\times$ $10^{-4}$   &        0.17   &   22.6  $\times$ $10^{-3}$   &      0.00    &    3.8  $\times$ $10^{-4}$  &      0.20  \\
NGC1872       &    1.2  $\times$ $10^{-4}$   &        0.96   &    3.6  $\times$ $10^{-3}$   &      0.10    &   10.5  $\times$ $10^{-4}$  &      0.19  \\
NGC2108       &   45.3  $\times$ $10^{-4}$   &        0.00   &    2.7  $\times$ $10^{-3}$   &      0.27    &    0.5  $\times$ $10^{-4}$  &      0.96  \\
NGC2203       &    1.9  $\times$ $10^{-4}$   &        0.37   &    5.2  $\times$ $10^{-3}$   &      0.00    &    0.3  $\times$ $10^{-4}$  &      0.80  \\
NGC2213       &    4.0  $\times$ $10^{-4}$   &        0.69   &   20.2  $\times$ $10^{-3}$   &      0.00    &    1.6  $\times$ $10^{-4}$  &      0.30  \\
\hline
\hline
\end{tabular*}
\end{table*}

\begin{figure}[ht]
\centering
\includegraphics[width=8cm]{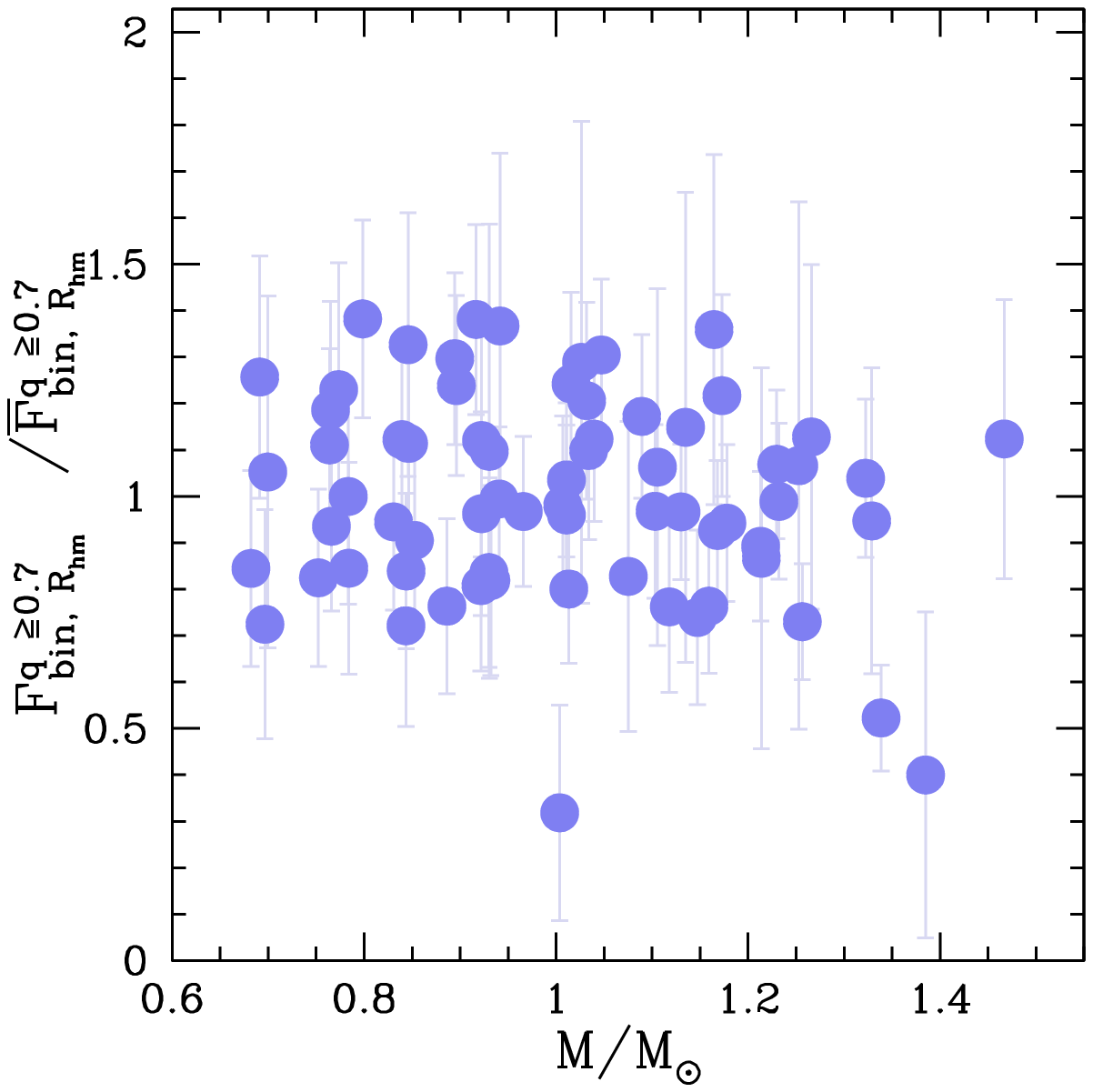}
\caption{Mass distribution of binaries. Binary fraction with mass ratios larger than 0.7 as a function of the mass of the primary star for all studied clusters.  
The binary fraction is normalized to the average binary fraction in a mass bin. 
}
\label{fig:mass2}
\end{figure}

To investigate the binary fraction in different intervals of primary-star mass, we divided the studied magnitude interval into five magnitude bins of equal width \footnote{ NGC\,1846 and NGC\,1872 are remarkable exceptions. Indeed, due to the smaller magnitude interval that we analyzed, we used four bins alone}. To do this, we used the mass-luminosity relations inferred from the best-fitting isochrones.
We estimated the fraction of binaries with q $\geq$ 0.7 and $r \leq R_{hm}$ in each bin and plotted this quantity as a function of the mean mass of the primary stars in that bin, as illustrated in Figure\,\ref{fig:mass}. 
A visual inspection of this figure suggests that the behavior of $F^{\rm q \geq 0.7}_{\rm bin, R_{hm}}$ as a function of the stellar mass varies from cluster to cluster.

Furthermore, for each cluster, we derived the $\chi^2$ value \footnote{We calculated the $\chi^{2}$ values according to the relation, $\chi^{2}=\sum_{0}^{N}(((Y-\bar{Y})^2)/N)$.} with respect to the average binary fraction, and the corresponding P-value, which indicates the deviation from a flat distribution. The latter is estimated by means of 1,000 Monte-Carlo simulations. In each simulation, we assumed a flat binary fraction corresponding to the observed average binary fraction and the same number of stars as in the observed CMD. 
The P-value is defined as the fraction of simulations with $\chi^2$ values greater than the observed $\chi^2$. The $\chi^2$ and P-values are listed in Table\,\ref{table:chi}.

We find that various clusters manifest a trend very close to flat distribution, as evident in NGC1806 ($P-value \sim $1). Conversely, NGC2108 ($P-value \sim $0) shows a decrease in binary fraction with increasing mass of the primary star.

To further compare the results from the different clusters, we calculated the average binary fraction ($\overline{F}^{\rm q \geq 0.7}_{\rm bin, R_{hm}}$) in the different mass bins, and plotted the $F^{\rm q \geq 0.7}_{\rm bin, R_{hm}}$/$\overline{F}^{\rm q \geq 0.7}_{\rm bin, R_{hm}}$ ratio as a function of the mean mass of the primary stars in that bin. The results from all the analysed clusters are plotted together in Figure \ref{fig:mass2}, where we observe that the general trend is that of a flat distribution, with a Spearman's rank correlation coefficient of $-$0.1. 


\subsection{Relations between the binary fraction and the mass ratio}

\begin{figure*}
\includegraphics[width=18cm]{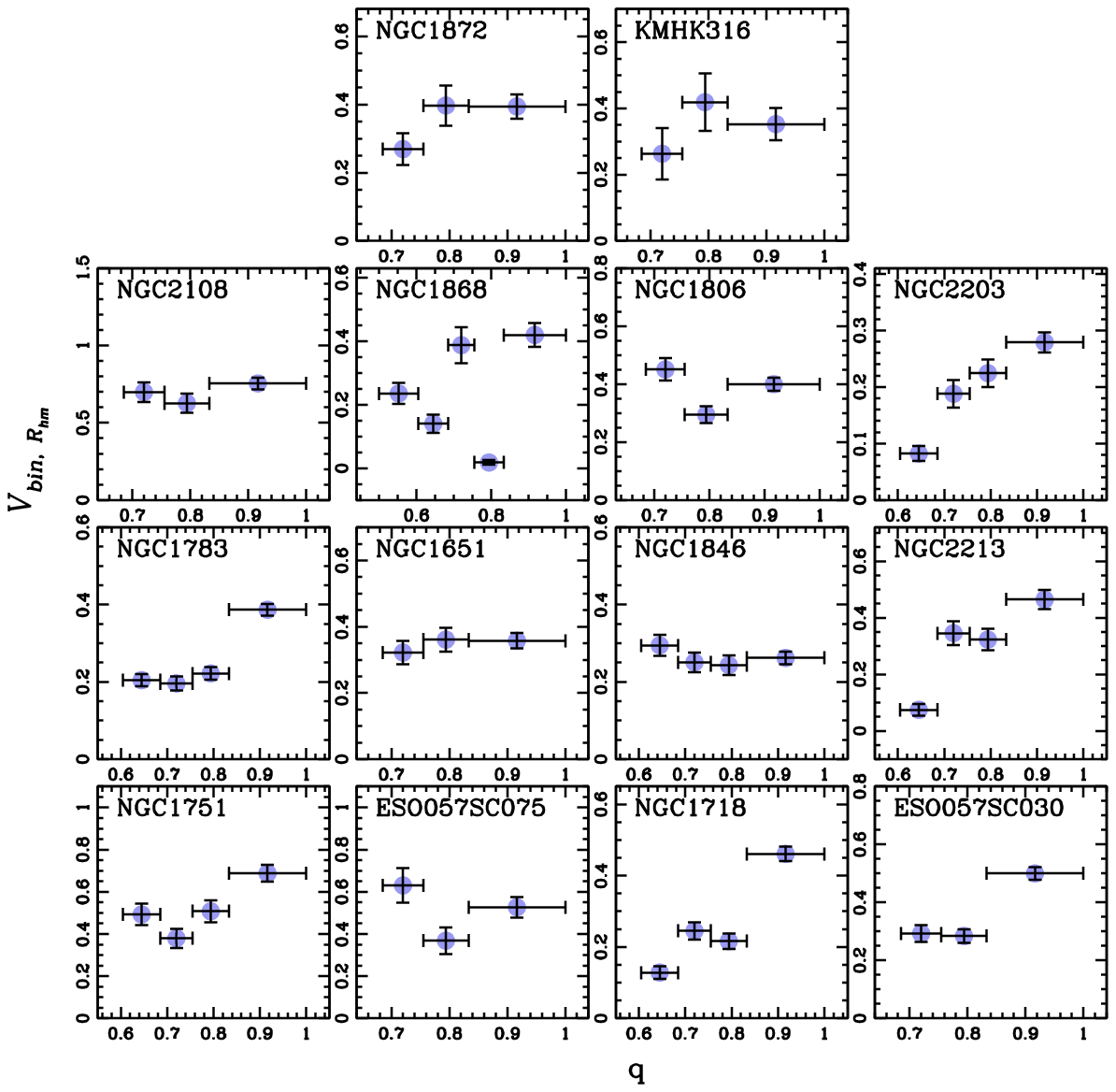}
\caption{Binary fraction and mass ratio parameter.
The frequency of binaries is plotted as a function of the mass ratio. The horizontal bars mark the mass-ratio intervals corresponding to each point. %
}
\label{fig:vbinq}
\end{figure*}

To constrain the relation between binary population and mass ratio, we calculated the binary fraction in different mass ratio bins and with $r \leq R_{hm}$. The q-intervals are selected within region B of the CMD with a prerequisite that the strips formed by the different $q$-loci should occupy the same area in the CMD. Therefore, we have selected loci of q values 0.6, 0.685, 0.755, 0.833, and 1.
Though the subsequent $q$ values are not equal in increment, they form equal-area strips in the CMD. 

We calculated the fraction of binaries in each mass ratio interval and derived the equivalent binary fraction, 
\begin{equation}
    V_{bin}=\frac{F_{bin}}{\Delta q}
\end{equation}
where $F_{bin}$ is the fraction of binaries in a given mass ratio interval, $\Delta q$.

The values of $V_{bin, R_{hm}}$ 
are plotted against $q$, as shown in Figure\,\ref{fig:vbinq}, whereas the $\chi^{2}$ and $P-values$ are provided in Table\,\ref{table:chi}.  We note that some clusters such as NGC1718, NGC2203, and NGC2213 (all with $P-value$ = 0) manifest an increase in the frequency of binaries with higher mass ratios,  while others, such as NGC\,1651 ($P-value$ $\geq$ 0.5) and NGC1846 ($P-value$ $\sim$ 0.5), exhibit a nearly flat distribution.
\begin{figure}[ht]
\centering
\includegraphics[width=8cm]{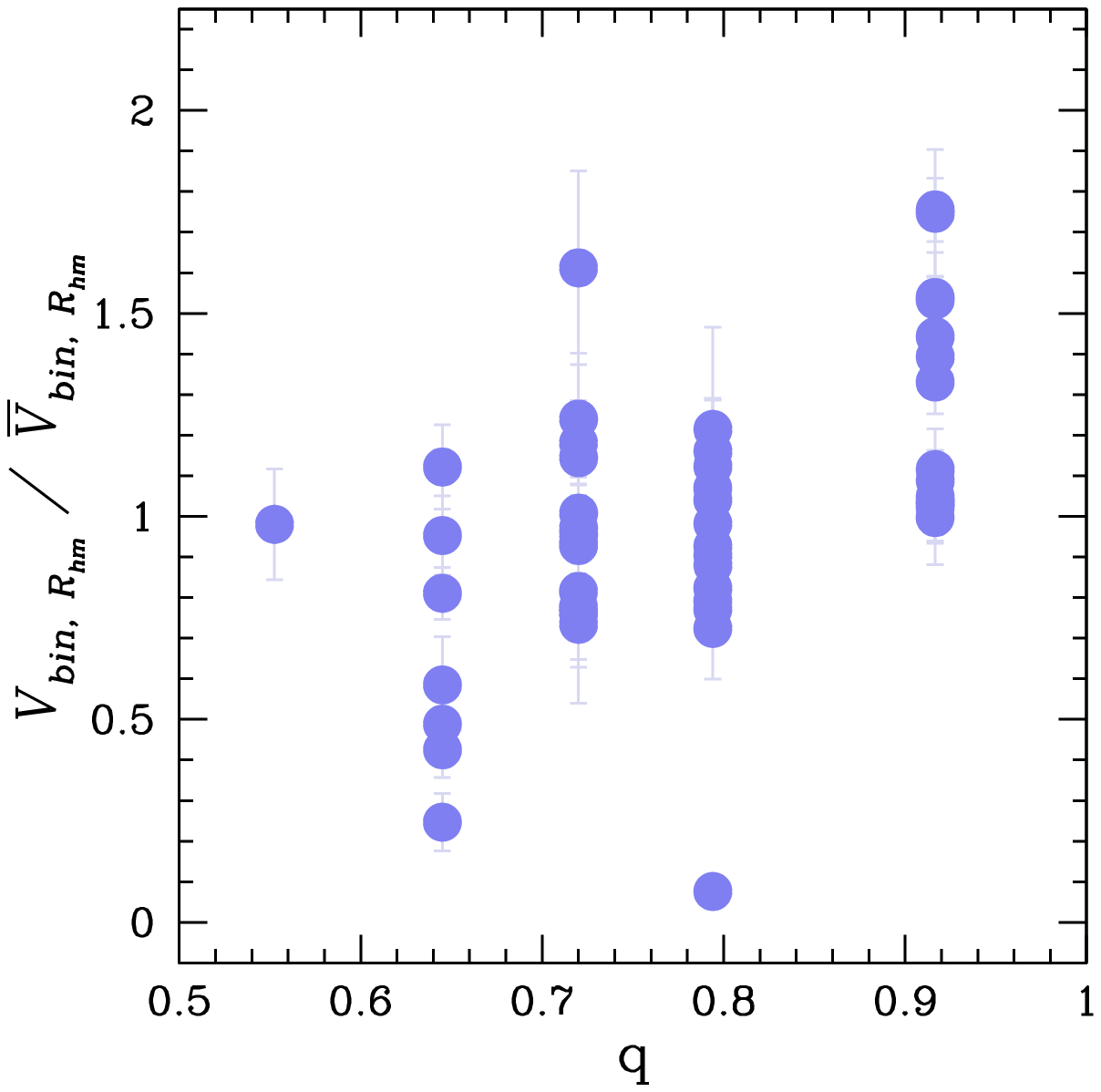}
\caption{Mass ratio distribution of binaries.
Frequency of binaries as a function of the mass ratio for all the studied clusters. 
}
\label{fig:vbinq11}
\end{figure}
\par By combining the results from all clusters, Figure \ref{fig:vbinq11} investigates the overall trend between binary fraction and mass ratio. To properly compare the different clusters, we have normalized the fraction of binaries by its average value. The distribution is nearly flat with a Spearman's rank correlation coefficient of $+$0.4.

\subsection{The radial distribution of binaries}

\begin{figure*}
\includegraphics[width=18cm]{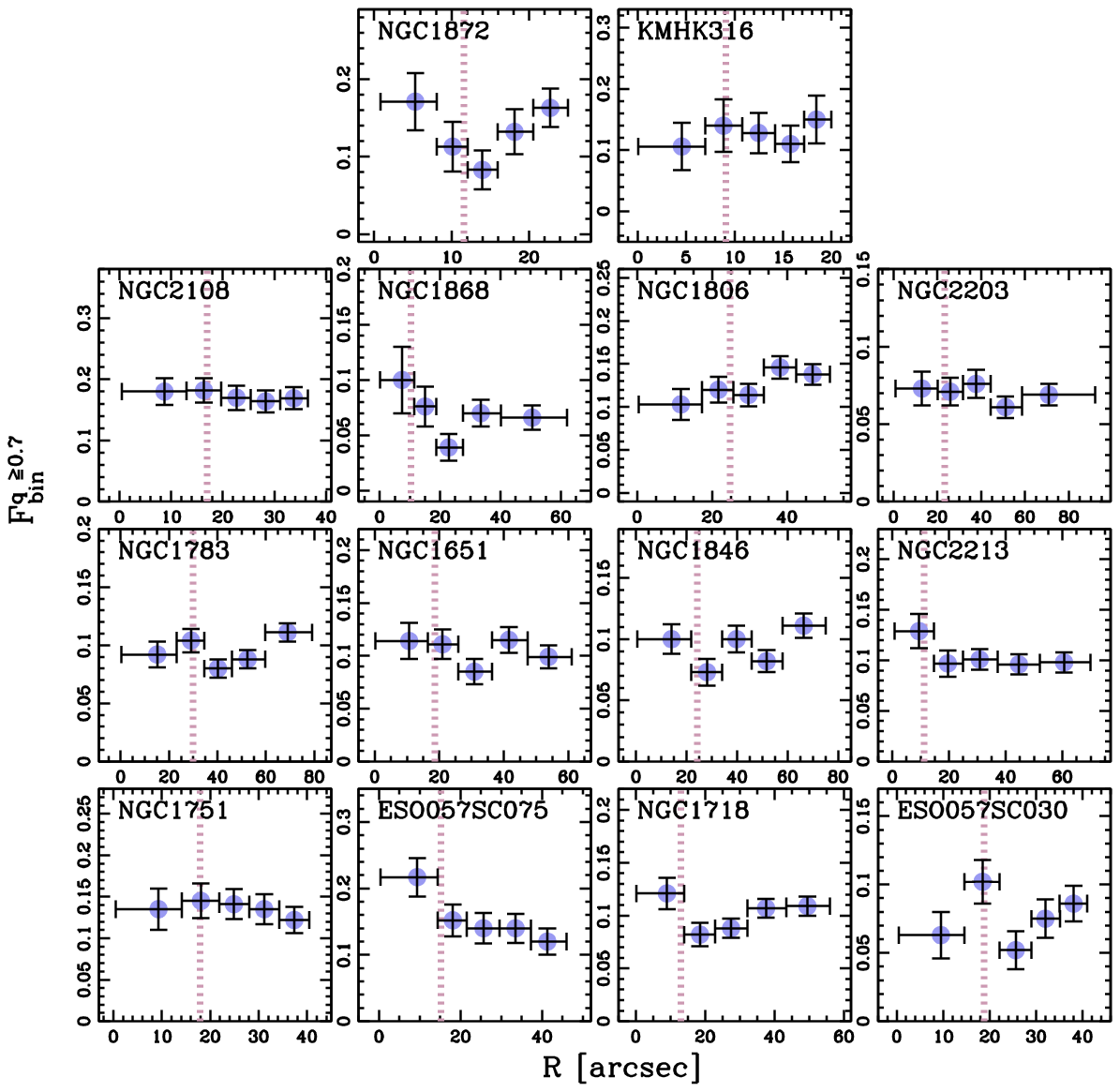}
\caption{Radial distribution of binaries. The core radii of the studied clusters are denoted with a dotted pink line.}
\label{fig:radial}
\end{figure*} 

\begin{figure*}[ht]
\begin{center}
    \includegraphics[trim={0 11.25cm 0 0},clip,width=18cm]{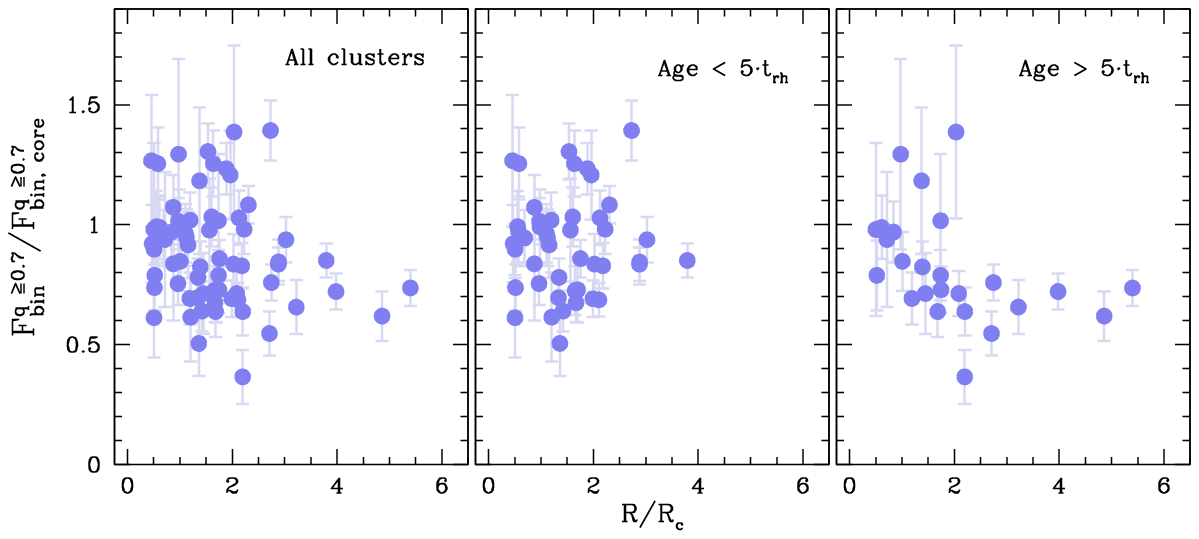}
\end{center}
\caption{Difference types of radial distribution. Binary fraction, normalized to the binary fraction in the core, as a function of the radial distance, in the unit of core radius.} 
\label{fig:radial1}
\end{figure*}

\par The radial distribution of binaries is examined by dividing the cluster region within the radial distance of $R_{2f}$ into five annuli, each with an equal number of MS stars. The binary fraction in each annulus is calculated using the procedure mentioned in Section\,\ref{sec:method} and is plotted as a function of the mean radial distance of the stars in that annulus from the cluster center, as in Figure \ref{fig:radial}. The corresponding $\chi^{2}$ and $P-values$ are reported in Table \ref{table:chi}.  

We find that in some clusters, such as ESO057SC075 and NGC\,1868, which are the studied clusters with a higher dynamical ages of 7 and 10 respectively, the binaries are centrally concentrated. However, the binaries of other clusters, including NGC\,1751 and NGC\,2108, show a flat distribution with $P-values \sim$ 1 whereas NGC1872 shows hints of a secondary peak. The change in radial distribution from flat to double peak, and then to a concentration in the center can be a function of dynamical age. Mass segregation is the main driver of this behaviour, as observed in the case of BSSs \citep{ferraro2012}.

The results for all clusters are plotted together in the left panel of Figure\,\ref{fig:radial1}, where we normalize the fraction of binaries to the fraction of binaries in the core, and express the radial distance in the units of the core radius.
We do not find evidence for a correlation between the fraction of binaries and the radial distance from the cluster center, as denoted by the correlation rank of $-$0.3. 
However, when we separate the clusters into two groups with ages either smaller or larger than 5 times their half-mass relaxation time, $t_{rh}$ (middle and right panels of Figure\,\ref{fig:radial1}, respectively), a pattern emerges. With a correlation rank of -0.6, most dynamically old clusters exhibit some hints of binary segregation into the centre, while the dynamically younger ones exhibit a flat distribution with a correlation rank of 0. 

\subsection{Candidate blue straggler stars}
\begin{figure}[ht]
\centering
\includegraphics[width=8.5cm]{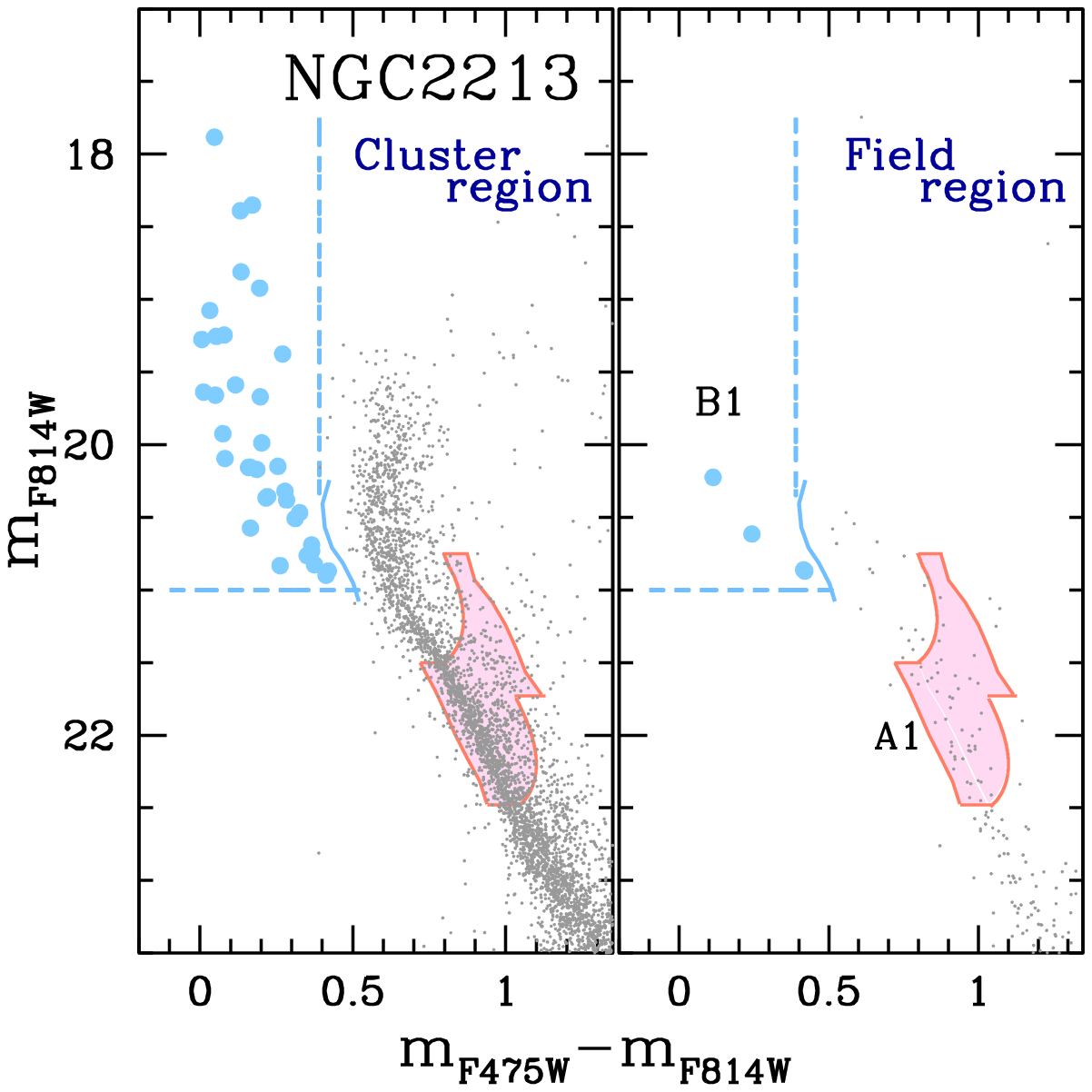}
\caption{Procedure to derive the BSS fraction.
$m_{\rm F814W}$ vs.\,$m_{\rm F475W}-m_{\rm F814W}$ CMD of stars in the cluster region of NGC\,2213 ({\it{left panel}}) and in the field region ({\it{right panel}}). The pink-shaded areas mark the CMD region A1, which is mostly populated by MS stars, while the stars in the CMD region B1 which hosts the candidate BSSs are marked with azure dots. The azure solid and dotted lines separate the B1 region from the remaining CMD area. See the text for details.
 }
\label{fig:bss}
\end{figure}

\begin{table*}
\caption{Fraction of  candidate BSSs for the studied clusters.}
\label{table:bss}
\centering
\begin{tabular*}{0.6\textwidth}{@{\extracolsep{\fill}}  c c c c }
\hline
\hline
    Cluster & $F_{BSS, R_{hm}}$ & $F_{BSS,core}$ & A$^{+}$ \\
    ID &  &  &  \\
\\
\hline
ESO057SC075  & 0.010  $\pm$  0.005    &  0.012   $\pm$ 0.006   &     0.12   $\pm$     0.00   \\
ESO057SC030  & 0.006  $\pm$  0.006    &  0.017   $\pm$ 0.008   &     0.02   $\pm$     0.01   \\
KMHK316      & 0.007  $\pm$  0.009    &  0.018   $\pm$ 0.015   &     0.16   $\pm$     0.04   \\
NGC1651      & 0.009  $\pm$  0.003    &  0.014   $\pm$ 0.006   &     0.25   $\pm$     0.03   \\
NGC1718      & 0.012  $\pm$  0.004    &  0.019   $\pm$ 0.006   &     0.07   $\pm$     0.02   \\
NGC1751      & 0.012  $\pm$  0.007    &  0.005   $\pm$ 0.009   &     0.04   $\pm$     0.02   \\
NGC1783      & 0.016  $\pm$  0.003    &  0.016   $\pm$ 0.003   &    -0.16   $\pm$     0.01   \\
NGC1806      & 0.021  $\pm$  0.003    &  0.019   $\pm$ 0.004   &     0.06   $\pm$     0.01   \\
NGC1846      & 0.018  $\pm$  0.004    &  0.026   $\pm$ 0.007   &     0.06   $\pm$     0.01   \\
NGC1868      & 0.006  $\pm$  0.003    &  0.018   $\pm$ 0.008   &     0.19   $\pm$     0.07   \\
NGC1872      & 0.000  $\pm$  0.002    &  0.001   $\pm$ 0.002   &    -0.07   $\pm$     0.01   \\
NGC2108      & 0.001  $\pm$  0.004    &  0.001   $\pm$ 0.005   &     0.15   $\pm$     0.03   \\
NGC2203      & 0.008  $\pm$  0.002    &  0.005   $\pm$ 0.003   &  0.00   $\pm$  0.01   \\
NGC2213      & 0.021  $\pm$  0.006    &  0.040   $\pm$ 0.012   &     0.30   $\pm$     0.03   \\
\hline
\hline
    \end{tabular*}
\end{table*}

\begin{figure}[ht]
\begin{center}
    \includegraphics[trim= 0 1cm 0 0, width=8.7cm]{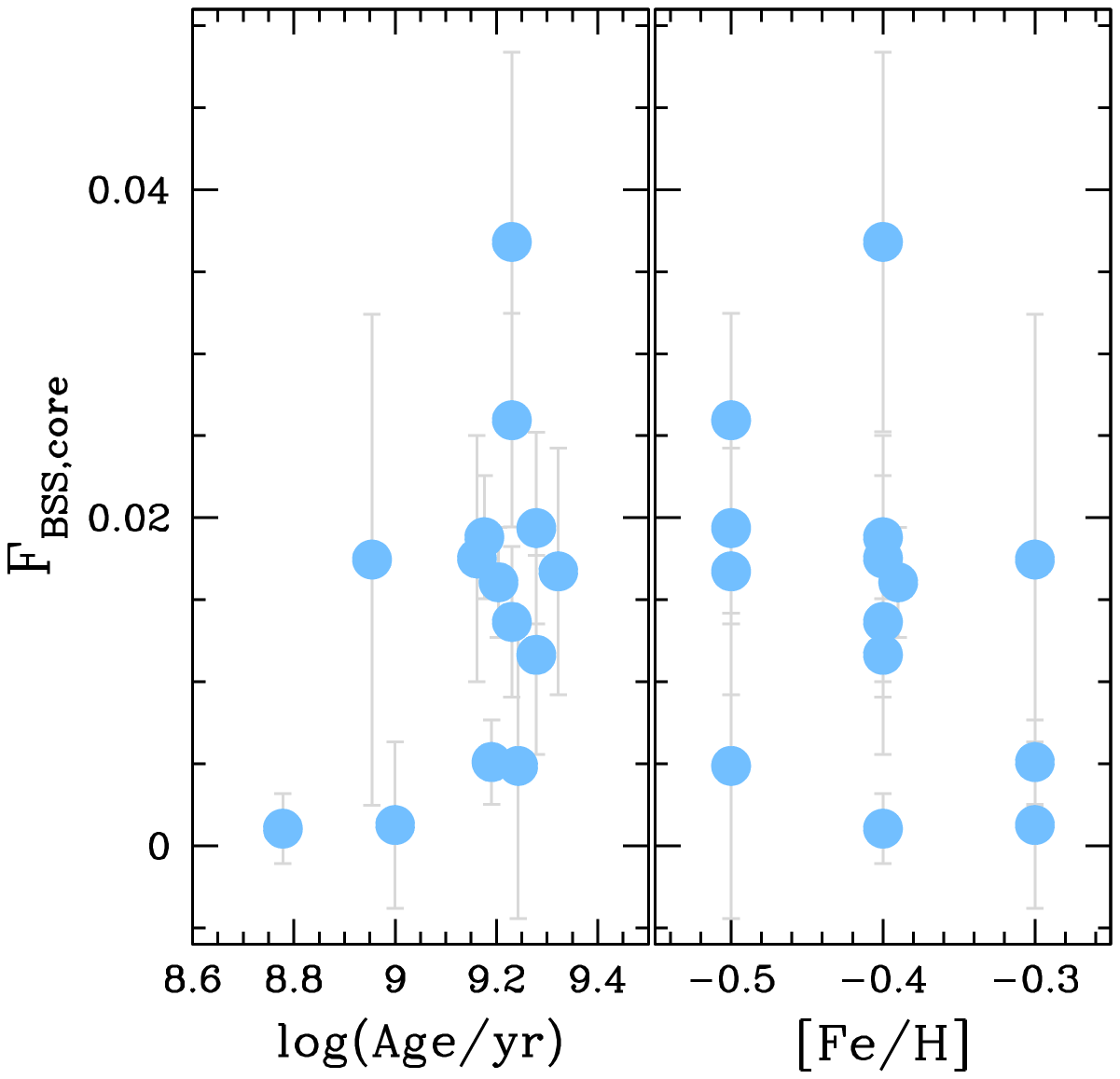}
\end{center}
\caption{BSS fraction and cluster parameters. Fraction of candidate BSSs in the core as a function of cluster age ({\it{left panel}}) and iron abundance ({\it{right panel}}). }
\label{fig:bssagemet}
\end{figure}

\begin{figure}[ht]
\centering
\includegraphics[width=9cm]{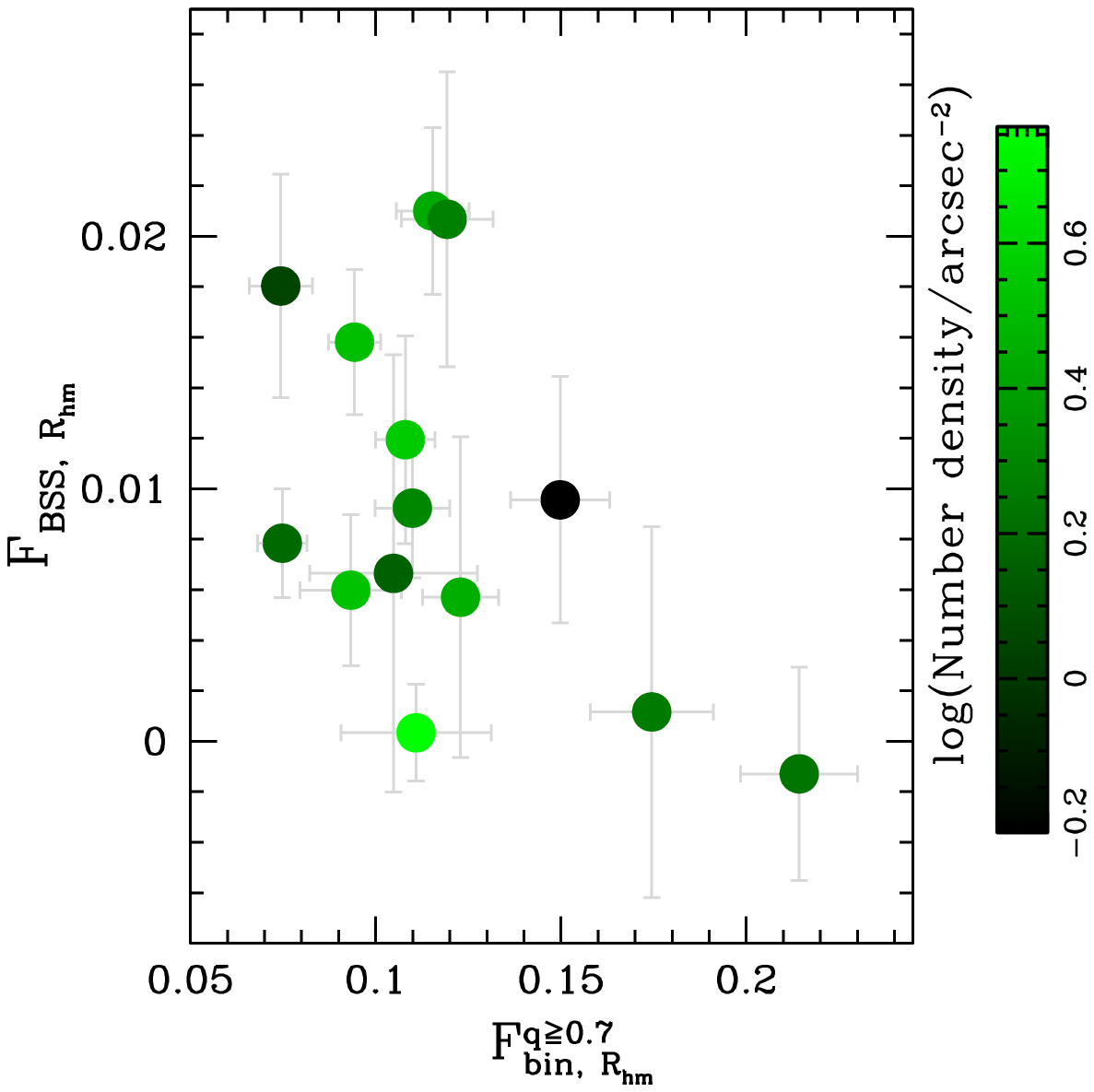}
\caption{Binary fraction and BSS fraction. Fraction of candidate BSSs as a function of the fraction of binaries with mass ratio, q$ \geq 0.7$ in the studied clusters. The clusters are colour-coded based on their density as shown in the colour-bar. }
\label{fig:bssfbinden}
\end{figure}

\par 
The CMDs of all analysed clusters show a sequence of stars in the blue and brighter side of the MSTO. This region of the CMD is populated by the BSSs, whose origin is traditionally associated with stellar mergers or the evolution of binary stars \citep[e.g.\,][and references therein]{sun2018}. Recent works argue that the sequences of blue stars observed in some intermediate-age LMC star clusters correspond to young stellar populations that emerged from the merging events in the Magellanic Clouds \citep{li2016a, hong2017}. Understanding the origin of the stars which we call candidate BSSs hereafter is beyond the purposes of our study where we determine some of their observational properties.

\par To derive the fraction of candidate BSSs in each cluster, we defined two regions in their CMD, namely A1 and B1, as illustrated in Figure\,\ref{fig:bss} for the cluster NGC\,2213. The A1 region (pink-shaded area) mostly hosts MS stars. It is similar to the region A of the CMD introduced in Section\,\ref{sec:method}, but spans an interval of one magnitude, in the filter F814W, along the fiducial line. The B1 region corresponds to the portion of the CMD that hosts the candidate BSSs and is limited by the azure solid and dotted lines. The continuous line corresponds to the MSFL shifted by four times $\sigma$ to the blue side, where $\sigma$ is the color uncertainty. The vertical line has the same color as the MSTO, whereas the horizontal line is 0.25 mag fainter than the brightest limit of region A1.

The fraction of candidate BSSs is calculated as,
\begin{equation}
\centering
    F_{BSS}=\frac{N^{B1}_{cluster} - N^{B1}_{field} }{N^{A1}_{cluster}-N^{A1}_{field}}
\end{equation}
where $N^{A1}_{cluster}$ and $N^{B1}_{cluster}$ are the numbers of stars, corrected for completeness, in the regions A1 and B1 of the cluster-region CMD, whereas $N^{A1}_{field}$ and $N^{B1}_{field}$ are the corresponding quantities for the field-region CMD that have been normalised by the ratio between the areas of cluster-region and the field-region.

The results are listed in Table\,\ref{table:bss} and reveal that the fraction of candidate BSSs within $R_{hm}$ ranges from 0.1\%, in NGC\,2108, to about 2\% in NGC\,1806 and NGC2213. Figure\, \ref{fig:bssagemet} reveals that there is no correlation between the fraction of candidate BSSs in the core and the cluster age or metallicity, as demonstrated by the Spearman's rank correlation coefficients of 0.3 and $-$0.3, respectively. Nevertheless, we confirm that clusters younger than $\sim$ 1 Gyr are unlikely to host BSSs \citep{rain2021a, cordoni2023}.

To further investigate the BSSs of the analyzed clusters, we derived the $A^{+}$ parameter which quantifies the difference between the cumulative distribution of the BSSs, $\phi_{BSS}$, and that of a reference population of stars with smaller masses, $\phi_{REF}$ \citep[e.g.][]{lanzoni2016}. 

The $A^{+}$ parameter is defined as:
\begin{equation}
    A^{+}(x)=\int_{x_{min}}^{x} \phi_{BSS} (x') dx' - \int_{x_{min}}^{x} \phi_{REF} (x') dx'
\end{equation}
where, $x=log(R/R_{hm})$ and $x_{min}$ is the minimum value that we sampled. We limited the analysis to stars within the half-mass radius and adopted the MS stars in the region A1 of the CMD as the reference population, as illustrated in Figure\,\ref{fig:bss} for the cluster NGC\,2213.
To account for the effect of field stars on the determination of A$^{+}$, we adopted a method that is based on the distributions of 1,000 groups of BSSs and reference stars.
 
To define each group, we first randomly associated each field-region star that populates the B1 and A1 portions of the CMD with a random position within the half-mass radius of the cluster. Then, we excluded them from the sample of BSSs and reference-population stars 
and derived a value of A$^{+}$ by using the remaining stars. The best determination of A$^{+}$ and the corresponding uncertainty are provided by the mean and r.m.s of the 1,000 determinations, respectively.
 
The results are listed in Table\,\ref{table:bss} and reveal that A$^{+}$ ranges from $\sim -0.16$ in NGC\,1783 to 0.30 in NGC\,2213 and that eleven out of fourteen clusters have positive A$^{+}$ values. Moreover, we find a mild correlation between $A^{+}$ and the cluster dynamical age that yields a Spearman's correlation rank of 0.6. This finding would provide constraints on the origin of the candidate BSSs \citep[e.g.][]{li2016a}.


\par The fraction of candidate BSSs are further plotted as a function of the binary fraction in Figure\,\ref{fig:bssfbinden}. 
We did not find a significant correlation between these two quantities, similar to what has been observed in Galactic open clusters \citep{cordoni2023}. Conversely, there is evidence that clusters with different densities seem to populate different sequences.

\subsection{Comparison with Galactic clusters}
\begin{figure*}
    \includegraphics[trim={0 11cm 0 0},width=18.5cm ]{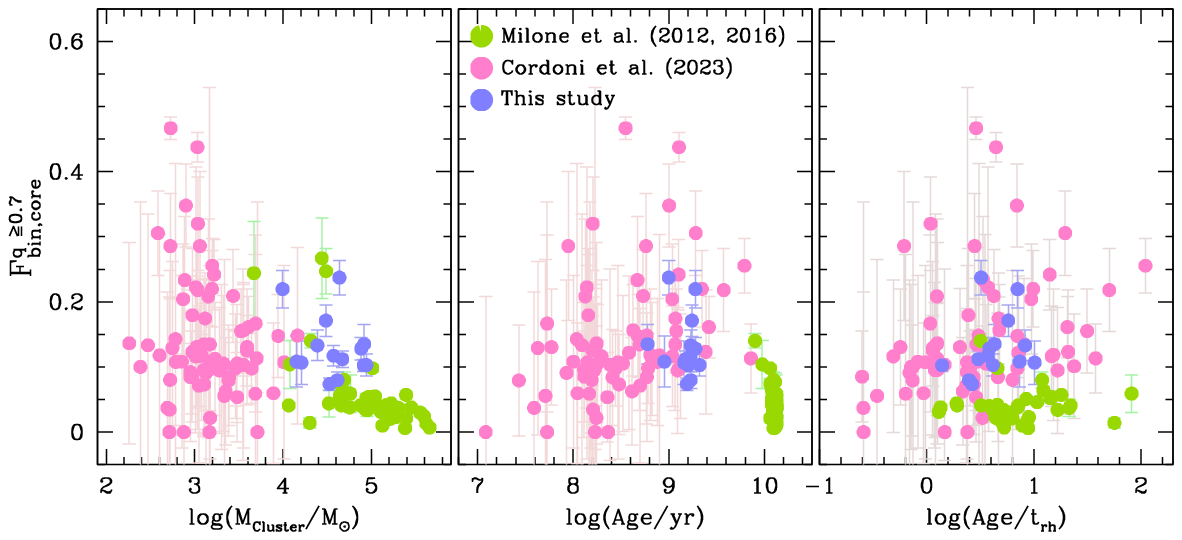}
\caption{Binary fraction in different environments. The plot compares the core binary fraction for clusters belonging to different environments. Galactic globular clusters, Galactic open clusters, and the Magellanic clouds globular clusters studied in this paper are denoted in lime green, magenta, and violet respectively. Binary fractions in these clusters are explored as a function of mass (\emph{left panel}), age (\emph{middle panel}), and dynamical age (\emph{right panel}) in logarithmic scale. }
\label{fig:fbinocgc}
\end{figure*}

\par To compare the binary fractions of Magellanic-Cloud star clusters and Galactic open and globular clusters, we combined the results of this work with results from the literature where binaries were analyzed homogeneously. The results are illustrated in Figure \ref{fig:fbinocgc} which compares the core binary fraction with $q \geq 0.7$ derived in this paper (purple dots) and those measured in Galactic GCs \citep[GGCs, green dots,][]{milone2012, milone2016} and Galactic open clusters \citep[pink dots,][]{cordoni2023}. The left panel of Figure\,\ref{fig:fbinocgc} shows the fraction of binary as a function of the cluster mass in the logarithmic scale. We used mass and $t_{rh}$ of Galactic GCs from \cite{baumgardt2018}, except for NGC\,6637, NGC\,6652, NGC\,6981, and Palomar\,1, for which the adopted values are from \cite{mclaughlin2005}. The Magellanic Cloud star clusters studied in our paper span a mass interval between 10$^{4}$ and 10$^{5}$ solar masses, which are poorly populated by the clusters studied in the previous studies.
  
\par When considering all clusters together, we find an anticorrelation between the binary fraction in the core and the cluster mass. However, for a fixed cluster mass, the binary fraction spans a wide range of values. In particular, the Magellanic cloud clusters studied in this paper exhibit a larger fraction of binaries than the bulk of GGCs with similar masses. Conversely, as shown in the middle and right panel of Figure\,\ref{fig:fbinocgc}, there is no evidence for a correlation between the fraction of binaries and the cluster age or the ratio between cluster age and $t_{rh}$.

\section{Summary and conclusions}\label{sec:conclusions}

\par We used data collected with the UVIS/WFC3 and ACS/WFC cameras on board {\it HST} to investigate the population of binaries and candidate BSSs along the MS of 14 Magellanic Cloud star clusters. Moreover, we determined the structure parameters of the star clusters, including the core radius and the central density, and estimated the mass function and the total mass of the cluster by fitting the density profile with an EEF profile.

The main results on binaries can be summarized as follows,
\begin{itemize}
    \item When we analyze the entire FoV, the fraction of binaries with mass ratio, $q \geq 0.7$ and $r \leq R_{hm}$ ranges from $\sim$7\%, in NGC\,1846, to more than $\sim 20$\%, in NGC\,2108. These values correspond to a total binary fraction between $\sim$20\% and 70\%, by assuming a flat mass-ratio distribution. We obtain similar results for the binaries in the cluster cores.
    \item There is no correlation between the fraction of binaries and the fraction of candidate BSSs, similar to what has been observed in the case of Galactic open clusters \citep{cordoni2023}. 
    \item When we combine the results from all clusters, we find no evidence for correlations between the fraction of binaries and either the mass of the primary star or the mass ratio. However, we notice various remarkable exceptions. As an example, the binary fraction decreases towards large stellar masses in NGC\,2108, while NGC\,1806 follows a flat distribution. Moreover, NGC\,1718, NGC\,2203, and NGC\,2213 exhibit a predominance of binaries with large mass ratios, whereas the fractions in NGC1651 and NGC1846 are distributed homogeneously.
    \item There is no evidence for significant differences in the radial distribution of binaries in most studied clusters. However, in clusters that are significantly older in comparison to their $t_{rh}$, the binaries are more centrally concentrated than single stars.
\end{itemize}

\par We have combined results on Magellanic Cloud clusters with those by a recent paper where we have performed a similar analysis on 78 Galactic open clusters \citep{cordoni2023}, and with results on 67 GGCs \citep{milone2012, milone2016}. In total, binaries have been now homogeneously studied in 159 star clusters.

\par The fraction of binaries does not correlate with either the cluster age or with the dynamical age. Conversely, we find a significant anti-correlation between the fraction of binaries in the core and the mass of the host cluster. However, clusters with similar masses exhibit a range of binary fractions that is wider than the observational errors. As an example, the star clusters studied in this paper typically host higher binary fractions than the GGCs with similar masses. This fact indicates that at least another parameter, in addition to cluster mass, determines the fraction of binaries in the core.

\section{Acknowledgement}
This work has received support from the European Research Council (ERC), under the European Union’s Horizon 2020 research innovation program (Grant Agreement ERC-StG 2016, No:716082, GALFOR, PI: Milone, $http://progetti.dfa.unipd.it/GALFOR$. AFM, GC, and APM acknowledge the support from the INAF-GTO-GRANTS 2022 (“Understanding the formation of globular clusters with their multiple stellar generations”, PI.A.F. Marino). SJ acknowledges support from the NRF of Korea (2022R1A2C3002992, 2022R1A6A1A03053472). TZ has received funding from the European Union’s Horizon 2020 research and innovation program under the MarieSklodowska-Curie Grant Agreement No. 101034319 and from the European Union – NextGenerationEU, beneficiary: Ziliotto. This study has used HST archival data from the Space Telescope Science Institute (STScI). We thank the anonymous referees for their constructive suggestions.

\footnotesize
\bibliographystyle{aa} 
\bibliography{aanda} 


\end{document}